\documentclass[sigconf]{acmart}
\AtBeginDocument{%
  }

\setcopyright{acmlicensed}

\usepackage{wrapfig}  
\usepackage{graphicx}
\usepackage{subcaption}
\usepackage{caption}
\usepackage{tipa}
\usepackage{blindtext}
\usepackage{lineno}
\usepackage{colortbl}

\usepackage{amsmath,amssymb,amsfonts}
\usepackage{amsthm}
\usepackage{graphicx}
\usepackage{multirow}
\usepackage{adjustbox}
\usepackage{wasysym}
\usepackage{footnote}
\usepackage{booktabs}
\usepackage{utfsym}
\usepackage{float}
\usepackage[dvipsnames]{xcolor}
\usepackage{multirow}
\usepackage{colortbl}
\usepackage{soul}

\usepackage{algorithm}

\usepackage{hyperref}
\usepackage{url}
\usepackage[dvipsnames]{xcolor}
\usepackage[table]{xcolor}  
\usepackage{pifont}         
\definecolor{mydarkblue}{rgb}{0,0.08,0.45}
\definecolor{darkgreen}{rgb}{0.0, 0.5, 0.0} 
\definecolor{myblue}{RGB}{235,235,250}
\definecolor{lightblue}{RGB}{225, 235, 250}
\definecolor{lightblue}{RGB}{230, 235, 245}  
\definecolor{lightgray}{RGB}{240, 240, 240}  
\definecolor{darkgray}{RGB}{220, 220, 220} 
\definecolor{superlightred}{rgb}{0.99, 0.92, 0.92}
\definecolor{darkgreen}{RGB}{50,100,0}
\definecolor{darkred}{RGB}{230, 150, 170}
\usepackage[most]{tcolorbox}

\usepackage{pgfplots}
\pgfplotsset{compat=1.17}

\usepackage{algorithm}
\usepackage[noend]{algpseudocode}
\usepackage{amsmath,amssymb}
\usepackage{xcolor}
\algrenewcommand\algorithmiccomment[1]{\hfill \textcolor{gray}{// #1}}

\definecolor{uclablue}{RGB}{159, 195, 224}

\definecolor{uclagold}{RGB}{156, 172, 234}

\definecolor{grayred}{RGB}{30, 140, 224}

\usepackage{color, colortbl}

\usepackage[table]{xcolor}
\definecolor{cyan}{rgb}{0.573, 0.675, 0.878}
\definecolor{limegreen}{rgb}{0.675, 0.843, 0.557}

\definecolor{casegreen}{RGB}{117, 189, 67}

\newtcolorbox{ttcolorbox}[1][]{
  colframe=uclagold,
  colback=uclagold!5!white,
  title=#1,
  fonttitle=\bfseries\sffamily,
}

\usepackage[most]{tcolorbox}
\tcbuselibrary{breakable}

\newtcolorbox{remarkbox}[1][]{ 
  colback=uclagold!10,
  colframe=uclagold!80!black,
  coltitle=white,
  title=#1,
  fonttitle=\bfseries,
  colbacktitle=uclagold!80!black,
  left=6pt, right=6pt, top=6pt, bottom=6pt,
  breakable
}

\copyrightyear{2026}
\acmYear{2026}
\setcopyright{cc}
\setcctype{by}
\acmConference[WWW '26]{Proceedings of the ACM Web Conference 2026}{April 13--17, 2026}{Dubai, United Arab Emirates}
\acmBooktitle{Proceedings of the ACM Web Conference 2026 (WWW '26), April 13--17, 2026, Dubai, United Arab Emirates}
\acmPrice{}
\acmDOI{10.1145/3774904.3792276}
\acmISBN{979-8-4007-2307-0/2026/04}

\author{Zhongyu Yang}
\orcid{0009-0003-0190-8256}
\affiliation{%
 \institution{BCML, Heriot-Watt University}
 \streetaddress{Edinburgh Campus, Riccarton}
 \city{Edinburgh}
 \postcode{EH14 4AS}
 \country{UK}}
\email{zy4028@hw.ac.uk}

\author{Wei Pang}
\orcid{0000-0002-1761-6659}
\affiliation{%
 \institution{BCML, Heriot-Watt University}
 \streetaddress{Edinburgh Campus, Riccarton}
 \city{Edinburgh}
 \postcode{EH14 4AS}
 \country{UK}}
\email{w.pang@hw.ac.uk}

\author{Yingfang Yuan}
\orcid{0000-0002-8925-9267}
\authornote{Corresponding author}
\affiliation{%
 \institution{BCML, Heriot-Watt University}
 \streetaddress{Edinburgh Campus, Riccarton}
 \city{Edinburgh}
 \postcode{EH14 4AS}
 \country{UK}}
\email{y.yuan@hw.ac.uk}

\renewcommand{\shortauthors}{Yang et al.}

\title{$\text{X}^\text{R}$: Cross-Modal Agents for Composed Image Retrieval}



\begin{document}

\begin{abstract}
Retrieval is being redefined by agentic AI, demanding multimodal reasoning beyond conventional similarity-based paradigms. 
Composed Image Retrieval (CIR) exemplifies this shift as each query combines a reference image with textual modifications, requiring compositional understanding across modalities.
While embedding-based CIR methods have achieved progress, they remain narrow in perspective, capturing limited cross-modal cues and lacking semantic reasoning.
To address these limitations, we introduce $\text{X}^\text{R}$, a training-free multi-agent framework that reframes retrieval as a progressively coordinated reasoning process. 
It orchestrates three specialized types of agents: \textit{imagination agents} synthesize target representations through cross-modal generation, \textit{similarity agents} perform coarse filtering via hybrid matching, and \textit{question agents} verify factual consistency through targeted reasoning for fine filtering. 
Through progressive multi-agent coordination, $\text{X}^\text{R}$ iteratively refines retrieval to meet both semantic and visual query constraints, achieving up to a 38\% gain over strong training-free and training-based baselines on FashionIQ, CIRR, and CIRCO, while ablations show each agent is essential. Code is available: \url{https://01yzzyu.github.io/xr.github.io/}.
\end{abstract}

\begin{teaserfigure}
  \centering
  \vspace{-4mm}
  \includegraphics[width=0.9\textwidth]{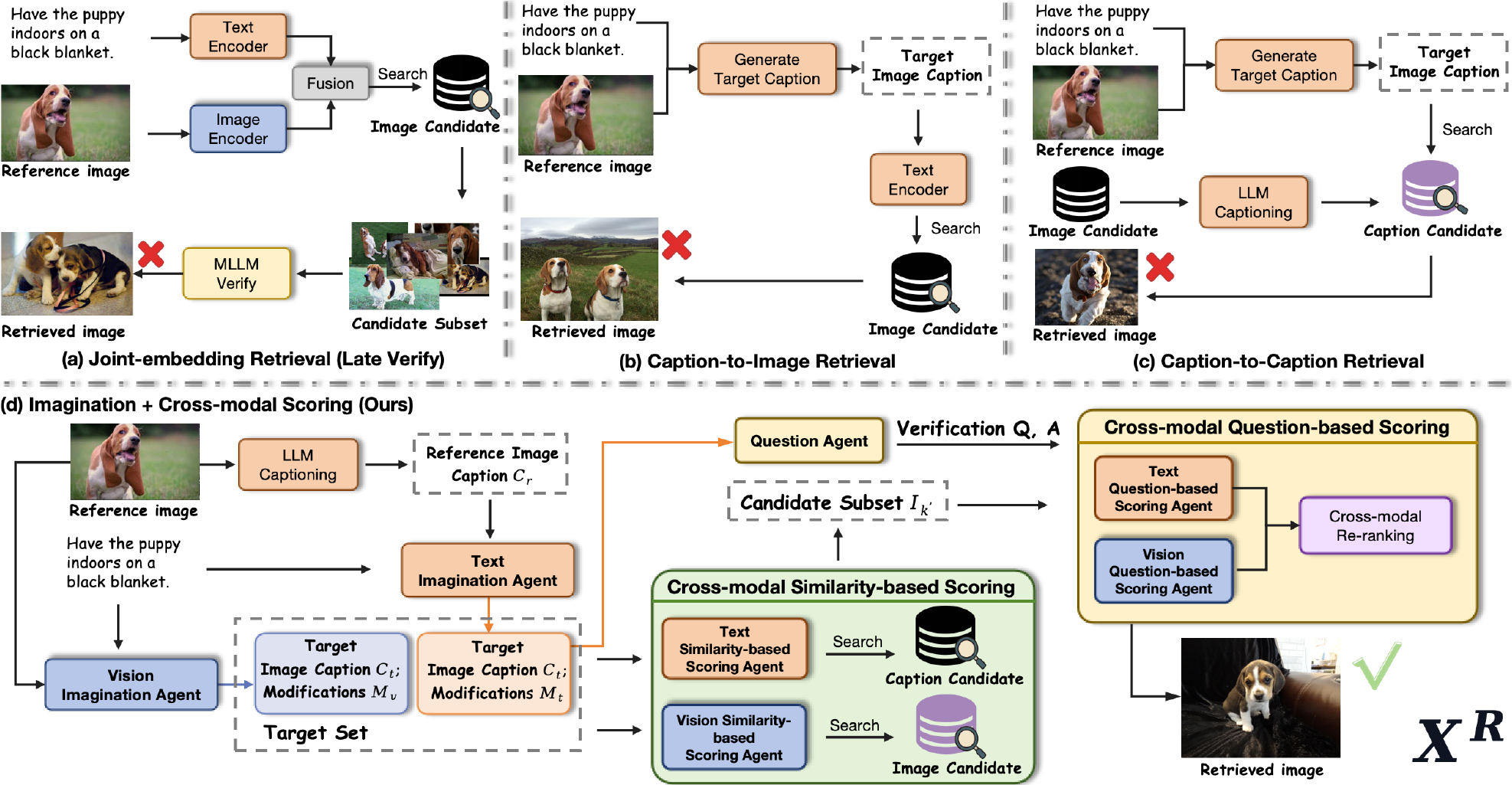}
  \vspace{-3mm}
  \caption{The workflows of existing CIR methods and ours: 
  (a) \textit{Joint embedding–based} methods encode a multimodal query into a shared space, but they often struggle to capture complex text-specified edits. 
  (b) \textit{Caption-to-Image} methods first generate a target caption from the multimodal query prior to retrieval, but they often fail to preserve fine-grained details.
  (c) \textit{Caption-to-Caption} methods build upon Caption-to-Image but restrict comparison to the text space, thereby discarding visual cues.
  (d) $\text{X}^\text{R}$ (ours) introduces an agentic AI framework with cross-modal agents and a progressive retrieval process consisting of an imagination stage followed by coarse-to-fine filtering, enabling robust reasoning that better aligns results with user intent.}
  \label{fig:teaser}
\end{teaserfigure}

\maketitle

\section{Introduction}
\label{sec:introduction}

Composed Image Retrieval (CIR)~\cite{vo2019composing, Hosseinzadeh2020ComposedQI, Chen2020ImageSW,
Liu2021ImageRO, Baldrati2023ZeroShotCI} is a retrieval paradigm where a query is explicitly composed by the user through a reference image and a modification text. 
CIR queries embody specific intent through the controlled composition of image and text.
This not only establishes CIR as a new direction in web information access, where users refine searches by combining images and text, but also links it to broader developments in retrieval-augmented agentic AI.
The demand for such interaction is evident in applications such as e-commerce~\cite{Chen2023UnifiedVR, zheng2023make} and search engines~\cite{xie2018people}, where navigating massive image repositories requires fine-grained multimodal control. 
Compared with conventional retrieval~\cite{Gudivada1995ContentBasedIR, Liu2007ASO, wan2014deep, Edge2024FromLT}, CIR is particularly challenging because it requires cross-modal reasoning to integrate heterogeneous signals rather than relying on a single unimodal cue. 
By pairing a reference image with textual modifications, CIR moves retrieval beyond simple content matching toward retrieving images that preserve reference semantics while faithfully applying the edits.

As illustrated in Figure~\ref{fig:teaser}, existing approaches can be broadly grouped into three categories: 
\textbf{(a) Joint embedding-based} methods project the multimodal query into a vector space and formulate CIR as similarity-based matching;
\textbf{(b) Caption-to-Image} methods first generate a target caption based on the multimodal query, then embed it and compare it with candidate image embeddings in terms of similarity; and
\textbf{(c) Caption-to-Caption} methods that directly compare candidate captions with the target caption.
Despite notable progress, these approaches exhibit persistent limitations. First, joint embeddings struggle to capture fine-grained, edit-specific correspondences due to imperfect cross-modal alignment. Second, a single similarity-based matching approach may fail to capture both textual and visual evidence. Third, the absence of cross-modal verification-based refinement undermines reliability, as each modality provides important information.


These challenges highlight our motivation that effective CIR must fully exploit cross-modal interactions. To address these challenges, we propose $\text{X}^\text{R}$, a training-free multi-agent framework that explicitly orchestrates cross-modal reasoning, providing robust retrieval under heterogeneous signals.
$\text{X}^\text{R}$ consists of three sequential modules: imagination, coarse filtering, and fine filtering. 
In imagination, agents construct a target proxy by generating captions from two cross-modal pairings, namely modification text with the reference image caption and modification text with the reference image, which helps reduce modality gaps and anchor the target semantics.
In coarse filtering, similarity-based agents evaluate candidates by producing multi-perspective scores using visual and textual cues, each conditioned on cross-modal captions. Reciprocal Rank Fusion (RRF) then aggregates these scores to form an initial ranked subset that addresses the limitations of single-criterion matching.
In fine filtering, question-based agents re-evaluate this subset through cross-modal factual verification by testing candidate images and captions with predicate-style queries, which mimic how humans validate retrieval consistency. Finally, verification scores are integrated with similarity scores through re-ranking to produce the final retrieval set, benefiting from both similarity-based matching and factual verification.

The similarity-based and question-based agents play complementary roles, where the former enables efficient high-level retrieval for broad coverage, while the latter enforces factual verification to refine results for accuracy. This design preserves diverse sources of evidence that single-score pipelines would otherwise overlook. Moreover, the cross-modality employed in both agents within $\text{X}^\text{R}$ enhances reliability by providing multi-perspective evidence. This is achieved through a combination of implicit coupling and explicit decoupling of modalities, enabling effective integration while maintaining per-modality interpretability. The proposed framework is tailored for edit-sensitive compositionality, capturing fine-grained modifications beyond the capability of unimodal systems.

We evaluate $\text{X}^\text{R}$ on three CIR benchmarks, CIRR, CIRCO, and FashionIQ, covering diverse retrieval scenarios from controlled reference-based queries to open-domain compositional settings. 
Across all datasets, $\text{X}^\text{R}$ consistently improves edit-sensitive retrieval accuracy over strong training-free and training-based baselines, demonstrating both its effectiveness and generality.
These results suggest practical value for web systems and applications, including personalized e-commerce search and multimodal recommendation.

In summary, our contributions are as follows:

\begin{itemize}
\setlength\itemsep{0pt}
\setlength\parsep{0pt}
\setlength\topsep{0pt}
    
\item We propose $\text{X}^\text{R}$, a training-free framework that orchestrates multiple cross-modal agents for CIR.  

\item We demonstrate the necessity of explicit cross-modality by showing its advantage over unimodal and single-score pipelines, which fail on edit-sensitive reasoning. 

\item Extensive experiments on CIRR, CIRCO, and FashionIQ show consistent gains over strong baselines, with ablations substantiating each module’s role, positioning $\text{X}^\text{R}$ as a general paradigm for multimodal retrieval.

\end{itemize}

\section{Related Works}
\label{relatedwork}

\noindent \textbf{Multimodal Agent Systems:}  
The rapid progress in MLLMs~\cite{openai2025-gpt5, wang2024qwen2, google2024gemini, guo2025deepseek} has enabled agentic frameworks with emerging abilities in autonomous planning, tool use, and decision-making. Such frameworks show strong potential for decomposing complex reasoning and coordinating across modalities~\citep{yao2023react, shinn2023reflexion, yang2025longvt, zhang2025openmmreasonerpushingfrontiersmultimodal, yang-etal-2025-mermaid}, though coordination remains fragile in practice.   
By iterative reflection and collaborative strategies, they mitigate hallucination and enhance interpretability, outperforming single-pass inference albeit at higher cost~\cite{Madaan2023SelfRefineIR, renze2024benefits, liu-etal-2025-instruct, inex}.
Yet most multimodal agents operate under a closed-world assumption, relying solely on internal inference and often hallucinating unsupported content~\cite{jiang2024multimodal, Durante2024AgentAS}, reflecting a lack of external grounding.  
In contrast, retrieval has long served as grounding in NLP pipelines, reducing uncertainty and improving adaptability~\cite{Liang2025ReasoningRV, singh2025agentic, agrawal2025gepareflectivepromptevolution}.  
Recent studies on retrieval-augmented agents,
such as Storm~\cite{shao2024assisting,costorm} and WikiAutoGen~\cite{yang2025wikiautogen} highlight the promise of retrieval-augmented agents, yet remain limited: Storm is text-centric, while WikiAutoGen extends to multimodality but in a narrow scope.
Overall, these findings underscore retrieval as a key enabler of reasoning, yet a systematic integration into general multimodal agents remains missing.

\noindent \textbf{Composed Image Retrieval:}  
CIR provides a natural testbed for retrieval-enhanced reasoning, where the task is to locate a target image given a reference image and textual modifications~\citep{song2025comprehensive, Zhang2025ComposedMR}.  
Most existing methods fuse features into a joint embedding and rank candidates by similarity, achieving coarse alignment but blurring fine-grained changes~\cite{vo2019composing, Chen2020ImageSW, li2024improving}.
Training-based models enhance representations but demand costly supervision and frequent retraining~\cite{huynh2025collm, xing2025context, Bao2025MLLMI2WHM, bai2024sentencelevel}.
Training-free approaches avoid task-specific supervision and generalize across domains. 
However, they rely on static fusion and one-shot pipelines, with little flexibility to refine uncertain retrieval results (e.g., candidate images or captions)~\cite{Cheng2025GenerativeTC, Li2025RethinkingPW, Li2024ImagineAS}.
Reasoning-style retrieval has been explored~\cite{tang2025reason, gu-etal-2025-toward, tu2025multimodal}, but existing methods remain fixed templates rather than adaptive workflows.
In practice, models still fail on fine-grained edits, for example, misinterpreting color changes in FashionIQ or mismatching object replacements in CIRR, underscoring the persistent limits of static similarity matching. 
This indicates that static similarity is not only brittle to edits but also fundamentally unable to capture compositional semantics.

In short, multimodal agents excel at reasoning but underexploit retrieval, while CIR methods leverage retrieval but lack reasoning, leaving the two largely disconnected. $\text{X}^\text{R}$ bridges this gap by embedding retrieval within an agentic workflow:
(1) imagination agents approximate the target image, preserving fine-grained details often missed by embeddings;
(2) similarity-based agents score candidates across modalities, reducing the rigidity of one-shot pipelines;
(3) question-based agents enforce factual checks, ensuring textual modifications are faithfully satisfied. Unified in a training-free system, these components elevate retrieval into dynamic reasoning, addressing the above limitations and yielding results that are both faithful to user intent and verifiable across modalities.
\section{Method}
\label{sec:method}

\begin{figure*}[t]
  \centering
  \includegraphics[width=\linewidth]{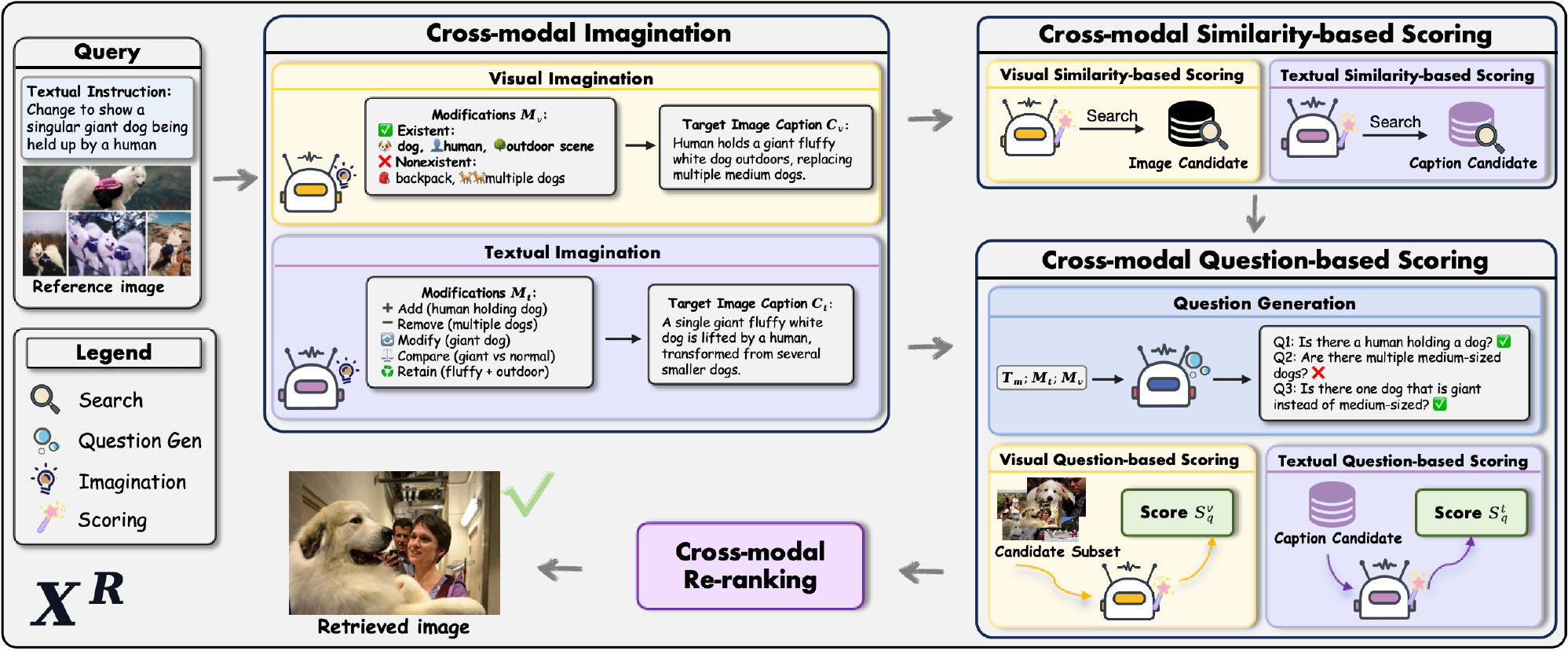}
  \vskip -2ex
  \caption{Framework of $\text{X}^\text{R}$. The multi-agent system integrates textual and visual imagination with cross-modal similarity and question-based scoring, followed by re-ranking. This multi-stage reasoning process exploits complementary cues from both modalities, effectively handling fine-grained modifications that single-modality approaches often miss.}
  \label{fig:framework}
\end{figure*}

\subsection{Preliminaries}
Given a multimodal query consisting of a reference image $I_{r}$ and a modification text $T_{m}$, CIR assumes the existence of an ideal target image $I_{i}$ that represents the desired outcome by preserving the visual characteristics of $I_{r}$ while incorporating the modifications specified by $T_{m}$. The CIR task then aims to retrieve a subset of images $\mathcal{I}^{\ast} \subseteq \mathcal{I}$, where $\mathcal{I} = \{ I_{1}, I_{2}, \ldots, I_{N} \}$ denotes the candidate image set containing $N$ images. Each $I \in \mathcal{I}^{\ast}$ is expected to approximate the ideal target image $I_{i}$.

To obtain $\mathcal{I}^{\ast}$, CIR typically proceeds in two stages. In the \emph{scoring} stage, each candidate image $I \in \mathcal{I}$ is evaluated against the query $(I_{r}, T_{m})$, which implicitly defines the ideal target image $I_{i}$. A matching score is then assigned, $S(I)=p(I \in \mathcal{I}^{\ast} \mid I_r, T_m)$, which represents the conditional probability that $I$ belongs to the target set $\mathcal{I}^{\ast}$ given the query. This score can also be viewed as a similarity measure between $I$ and the ideal target $I_{i}$. In the subsequent \emph{ranking} stage, candidates are ordered by their scores, and the top-$k$ images are selected to form $\mathcal{I}^{\ast}$, where $|\mathcal{I}^{\ast}| = k$.

\subsection{$\text{X}^\text{R}$ Framework}
To address CIR, we propose $\text{X}^\text{R}$, a training-free multi-agent framework that emphasizes the role of cross-modality in improving retrieval accuracy. Unlike existing approaches, our framework is composed of three cross-modal modules: imagination, coarse filtering, and fine filtering. In the coarse filtering stage, similarity-based scoring agents identify an initial subset of candidates that approximate the ideal target image through similarity evaluation. In the fine filtering stage, question-based scoring agents perform factual verification to further filter and refine this subset, producing the final ordered set $\mathcal{I}^*$. These two components complement each other, working together to progressively narrow down candidates and improve ranking. Throughout the process, cross-modal mechanisms offer multi-perspective support that enhances robustness and reliability in retrieval while reducing the risks associated with multimodal misalignment.

The workflow of $\text{X}^\text{R}$ is outlined in Algorithm~\ref{alg:exp} and depicted in Figure~\ref{fig:teaser}. The caption agent $\mathcal{A}_c$ generates a set of candidate captions $\mathcal{C}$ by iteratively producing a caption for each $I \in \mathcal{I}$ (Line~1). It also generates the caption $C_r$ for the reference image $I_r$ (Line~2). To construct or imagine the ideal target image $I_i$, the text imagination agent $\mathcal{A}_t^i$ and the vision imagination agent $\mathcal{A}_v^i$ jointly form the cross-modal imagination module. These two agents generate captions $C_t$ and $C_v$, respectively, to describe $I_i$ from different modalities. In addition, $\mathcal{A}_t^i$ produces $M_t$, which specifies the manipulations required to transform $C_r$ with $T_m$ in order to approximate $I_i$, thereby unifying the information in the text modality. Meanwhile, $M_v$ denotes a set indicating the presence or absence of visual attributes.

In coarse filtering, to evaluate each candidate image $I$, $\text{X}^\text{R}$ employs a text similarity-based scoring agent $\mathcal{A}_t^s$ (Line~6) and a vision similarity-based scoring agent $\mathcal{A}_v^s$ (Line~7). These agents assess the $a$-th candidate from different modalities, using $C_a$ and $I_a$ respectively, conditioned on $C_t$ and $C_v$. Each scoring agent produces two scores, denoted $s^t$ and $s^v$, through a process termed cross-modal multiperspective scoring, which operates within the embedding space based on similarity. The text-based score $s^t$ and vision-based score $s^v$ are then aggregated across agents in Lines~8 and~9. In Line~10, the score vectors $S^t = \{s_1^t, \dots, s_N^t\}$ and $S^v = \{s_1^v, \dots, s_N^v\}$ are processed separately using a Reciprocal Rank Fusion function. The resulting rank scores are combined, after which candidates are ranked and filtered to yield the top-$k^\prime$ results, denoted by $\mathcal{I}^{k^\prime}$.

To enhance the ranking accuracy of the selected top-$k^\prime$ candidates and further refine this subset, $\text{X}^\text{R}$ incorporates cross-modal question-based scoring agents in the fine filtering stage. The question agent $\mathcal{A}^q$ (Line 11) formulates a set of questions $Q$ with corresponding answers $A$ based on the information $M_t$, $M_v$, and $T_m$, focusing on the essential attributes that the ideal target image $I_i$ should contain. Unlike the similarity-based scoring agents, $\mathcal{A}^q$ emphasizes the critical differences between a candidate image $I$ and the ideal target $I_i$. Two question-based scoring agents, $\mathcal{A}_t^q$ and $\mathcal{A}_v^q$, then use $Q$ and $A$ to re-evaluate candidates in $\mathcal{I}^{k^\prime}$ from the text and vision modalities, applying $C_a$ and $I_a$ respectively (Lines 13–14). Finally, the scores from the question-based scoring agents are combined with the aggregated similarity-based scores through a cross-modal re-ranking function (Line 15). The candidates are then re-ordered based on the resulting scores to form the fine-filtered subset, which constitutes the final output $\mathcal{I}^*$ of size $k$, where $k < k^\prime$. The following paragraphs describe in detail the roles of imagination, coarse filtering, and fine filtering.

\algrenewcommand\algorithmicrequire{\textbf{Input:}}
\algrenewcommand\algorithmicensure{\textbf{Output:}}

\begin{algorithm}
\caption{$\text{X}^\text{R}$}
\label{alg:exp}
\begin{algorithmic}[1] 
\Require $\mathcal{I}$ cadidate image set, $I_r$ reference image, $T_m$ modification text
\Ensure $\mathcal{I}^{\ast}$ target image set
\newcommand{\LineComment}[1]{\Statex \textcolor{gray}{\# #1}}

\LineComment{\color{blue} initialization} 
\State $\mathcal{C} = \mathcal{A}_c(\mathcal{I})$ \Comment{\small candidate image caption set}
\State $C_r = \mathcal{A}_c(I_r)$

\LineComment{\color{blue} imagination}
\State $ M_t, C_{t} =\mathcal{A}_t^i (T_m, C_r)$ \Comment{\small text imagination agent}
\State $ M_v, C_{v} =\mathcal{A}_v^i (T_m, I_r)$ \Comment{\small vision imagination agent}

\LineComment{\color{blue} coarse filtering}
\For{$a = 1$ \textbf{to} $N$} \Comment{\small $I_a \in \mathcal{I}, C_a \in \mathcal{C}$}
\State $ s_t^t, s_t^v =\mathcal{A}_t^s ( C_{t}, C_{v}, C_a)$ \Comment{\small text similarity-based scoring agent}
\State $ s_v^t, s_v^v =\mathcal{A}_v^s (C_{t}, C_{v}, I_a)$ \Comment{\small vision similarity-based scoring agent}
\State $s^t_a = s_t^t + s_v^t $ 
\State $s^v_a = s_v^v + s_t^v $ 
\EndFor

\State $\mathcal{I}^{k^\prime}= \operatorname{ranking}(S^t, S^v)$ \Comment{\small reciprocal rank fusion function}

\LineComment{\color{blue}fine filtering}
\State $Q, A = \mathcal{A}^q (M_t, M_v, T_m)$  
\Comment{\small question agent}
\For{$a = 1$ \textbf{to} $k^\prime$} 
\State  $s^t_q = \mathcal{A}_t^q (C_a, Q, A)$ \Comment{\small text question-based scoring agent}
\State  $s^v_q = \mathcal{A}_v^q (I_a, Q, A)$\Comment{ \small vision question-based scoring agent}
\EndFor
\State $S^{k^\prime} \leftarrow (S^t_q + S^v_q) * \operatorname{norm}(\lambda S^t+(1-\lambda)S^v)$ \Comment{\small cross-modal re-ranking}

\State $\mathcal{I}^* = \operatorname{re-ranking}(S^{k^\prime})$ 
\\
\Return $\mathcal{I}^*$ \Comment{\small $|\mathcal{I}^*|=k$}

\end{algorithmic}
\end{algorithm}

\subsection{Imagination}
The imagination agents $\mathcal{A}_t^i$ and $\mathcal{A}_v^i$ play a central role in $\text{X}^\text{R}$. In CIR, accurate retrieval requires a clear representation of the ideal target image $I_i$ that satisfies the multimodal query, since defining $I_i$ provides prior knowledge and evidence to guide the retrieval of similar candidates. This prior knowledge is critical, as incorrect priors can trigger a cascade of errors. To address this, the agents $\mathcal{A}_t^i$ and $\mathcal{A}_v^i$ are designed to imagine and approximate the ideal target image by generating cross-modal captions that capture complementary aspects of $I_i$. The cross-modality arises from the fact that $\mathcal{A}_v^i$ and $\mathcal{A}_t^i$ take different inputs: $\mathcal{A}_v^i$ uses the reference image $I_r$ and $\mathcal{A}_t^i$ uses the reference caption $C_r$, and both are conditioned on the modification text $T_m$. Their outputs are the vision-based caption $C_v$ and the text-based caption $C_t$, respectively. 

\begin{table*}[]
\caption{Performance comparison on CIRCO and CIRR test set. The best results are in bold, and the second best are underlined.}
\vskip -3ex
\label{tab:cirr_circo}
\resizebox{\textwidth}{!}{
\begin{tabular}{c|llc|rrrr||rrrr|rrrr}
\bottomrule
\multicolumn{1}{c|}{\multirow{2}{*}[-0.5ex]{\textbf{Backbone}}} & 
\multicolumn{1}{c}{\multirow{2}{*}[-0.5ex]{\textbf{Method}}} & \multicolumn{1}{c}{\multirow{2}{*}{\textbf{Venue}}} & \multicolumn{1}{c|}{\multirow{2}{*}{\textbf{Training-free}}} & \multicolumn{4}{c||}{\textbf{CIRCO}} & \multicolumn{4}{c}{\textbf{CIRR}} & \multicolumn{3}{c}{CIRR$_{subset}$} \\
\cline{5-8}  \cline{9-12} \cline{13-15}
\multicolumn{1}{c|}{} & \multicolumn{1}{c}{} & \multicolumn{1}{c}{} & \multicolumn{1}{c|}{} & \multicolumn{1}{c}{mAP@5} & \multicolumn{1}{c}{mAP@10} & \multicolumn{1}{c}{mAP@25} & \multicolumn{1}{c||}{mAP@50} & \multicolumn{1}{c}{R@1} & \multicolumn{1}{c}{R@5} & \multicolumn{1}{c}{R@10} & \multicolumn{1}{c|}{R@50} & \multicolumn{1}{c}{R@1} & \multicolumn{1}{c}{R@2} & \multicolumn{1}{c}{R@3} \\
\hline
\multirow{9}{*}{\rotatebox{90}{CLIP-ViT-B/32}} & PALAVRA~\cite{Cohen2022ThisIM} & \textit{ECCV 2022} & \usym{1F5F6} & 4.61 & 5.32 & 6.33 & 6.80 & 16.62 & 43.49 & 58.51 & 83.95 & 41.61 & 65.30 & 80.95 \\
 & SEARLE~\cite{Baldrati2023ZeroShotCI} & \textit{ICCV 2023}  & \usym{1F5F6}  & 9.35 & 9.94 & 11.13 & 11.84 & 24.00 & 53.42 & 66.82 & 89.78 & 54.89 & 76.60 & 88.19 \\
 & SEARLE-OTI~\cite{Baldrati2023ZeroShotCI} & \textit{ICCV 2023}  & \usym{1F5F6}  & 7.14 & 7.38 & 8.99 & 9.60 & 24.27 & 53.25 & 66.10 & 88.84 & 54.10 & 75.81 & 87.33 \\
 & iSEARLE~\cite{iSEARLE} & \textit{TPAMI 2025} & \usym{1F5F6} & 10.58 & 11.24 & 12.51 & 13.26 & 25.23 & {55.69} & 68.05 & {90.82} & - & - & - \\
 & iSEARLE-OTI~\cite{iSEARLE} & \textit{TPAMI 2025} & \usym{1F5F6} & 10.31 & 10.94 & 12.27 & 13.01 & 26.19 & 55.18 & 68.05 & 90.65 & - & - & - \\
 & CIReVL~\cite{CIReVL} &\textit{ICLR 2024} & \usym{2714}& 14.94 & 15.42 & 17.00 & 17.82 & 23.94 & 52.51 & 66.00 & 86.95 & 60.17 & 80.05 & 90.19 \\
 & LDRE~\cite{yang2024ldre} & \textit{SIGIR 2024} & \usym{2714} & {17.96} & {18.32} & {20.21} & {21.11} & {25.69} & 55.13 & {69.04} & \underline{89.90} & {60.53} & {80.65} & {90.70} \\
 & ImageScope~\citep{luo2025imagescope} & \textit{WWW 2025} & \usym{2714} & \underline{22.36} & \underline{22.19} & \underline{23.03} & \underline{23.83} & \underline{34.36} & \underline{60.58} & \underline{71.40} & 88.41 & \underline{74.63} & \underline{87.93} & \underline{93.83} \\
 \rowcolor{lightblue}
  & \textbf{$\text{X}^\text{R}$(Ours)} & \textit{\textbf{Proposed}} & \usym{2714}   & \textbf{27.51} & \textbf{28.33} & \textbf{30.28} & \textbf{30.95} & \textbf{43.06} & \textbf{73.86} & \textbf{83.15} & 	\textbf{94.36} & \textbf{77.54} & \textbf{90.27} & \textbf{95.21} \\
 \hline
\multirow{11}{*}{\rotatebox{90}{CLIP-ViT-L/14}} & Pic2Word~\cite{saito2023pic2word} & \textit{CVPR 2023} & \usym{1F5F6} & 8.72 & 9.51 & 10.64 & 11.29 & 23.90 & 51.70 & 65.30 & 87.80 & - & - & - \\
 & SEARLE~\cite{Baldrati2023ZeroShotCI} & \textit{ICCV 2023}  & \usym{1F5F6} & 11.68 & 12.73 & 14.33 & 15.12 & 24.24 & 52.48 & 66.29 & 88.84 & 53.76 & 75.01 & 88.19 \\
 & SEARLE-OTI~\cite{Baldrati2023ZeroShotCI} & \textit{ICCV 2023}  & \usym{1F5F6}  & 10.18 & 11.03 & 12.72 & 13.67 & 24.87 & 52.32 & 66.29 & 88.58 & 53.80 & 74.31 & 86.94 \\
 & iSEARLE~\cite{iSEARLE} & \textit{TPAMI 2025} & \usym{1F5F6} & 12.50 & 13.61 & 15.36 & 16.25 & 25.28 & 54.00 & {66.72} & 88.80 & - & - & - \\
 & iSEARLE-OTI~\cite{iSEARLE} & \textit{TPAMI 2025} & \usym{1F5F6} & 11.31 & 12.67 & 14.46 & 15.34 & 25.40 & 54.05 & 67.47 & 88.92 & - & - & - \\
& LinCIR~\cite{LinCIR} & \textit{CVPR 2024} & \usym{1F5F6}& 12.59 & 13.58 & 15.00 & 15.85 & 25.04 & 53.25 & 66.68 & - & 57.11 & 77.37 & 88.89 \\
& FTI4CIR~\cite{Lin2024FinegrainedTI} &  \textit{SIGIR 2024} & \usym{1F5F6}& 15.05 & 16.32 & 18.06 & 19.05 & 25.90 & 55.61 & 67.66 & \underline{89.66} & 55.21 & 75.88 & 87.98 \\
& CIReVL~\cite{CIReVL} &\textit{ICLR 2024} & \usym{2714}& 18.57 & 19.01 & 20.89 & 21.80 & 24.55 & 52.31 & 64.92 & 86.34 & 59.54 & 79.88 & 89.69 \\
 & LDRE~\cite{yang2024ldre} & \textit{SIGIR 2024} & \usym{2714} & {23.35} & {24.03} & {26.44} & {27.50} & {26.53} & {55.57} & 67.54 & 88.50 & {60.43} & {80.31} & {89.90} \\
 & ImageScope~\citep{luo2025imagescope} & \textit{WWW 2025} & \usym{2714} & \underline{25.39} & \underline{25.82} & \underline{27.07} & \underline{27.98} & \underline{34.99} & \underline{61.35} & \underline{71.49} & 88.84 & \underline{74.94} & \underline{88.24} & \underline{94.02} \\
 \rowcolor{lightblue}
  & \textbf{$\text{X}^\text{R}$(Ours)} & \textit{\textit{\textbf{Proposed}}} & \usym{2714} & \textbf{31.38}  & \textbf{32.88}  & \textbf{35.46}  & \textbf{36.50}  & \textbf{43.13} &\textbf{73.59} & \textbf{83.09} & \textbf{94.05} & \textbf{77.98} & \textbf{90.68} & \textbf{95.06} \\
\toprule
\end{tabular}
}
\end{table*}

The design of cross-modal imagination is motivated by the observation that information from different modalities provides complementary strengths and weaknesses when estimating $I_i$. Specifically, the pair $(T_m, C_r)$ is more straightforward, which facilitates the extraction of key information for depicting $I_i$. In contrast, the pair $(T_m, I_r)$ combines textual and visual inputs, where the image contributes fine-grained details that complement the textual description. We argue that combining both perspectives enables the model to adapt flexibly to diverse situations encountered in real-world retrieval tasks. As shown in Figure~\ref{fig:framework}, in cross-modal imagination, $C_v$ and $C_t$ generate captions that are similar but not identical. The former captures fine-grained visual details, such as ``outdoors'' and ``multiple medium dogs'', which are grounded in the image. The latter, by contrast, emphasizes semantic transformation, for example ``transformed from several smaller dogs'', reflecting a more abstract and text-driven perspective rather than visually specific cues.

Additionally, $\mathcal{A}_t^i$ is designed to output $M_t$, which represents the modifications between $C_t$ and $C_r$. While $T_m$ is provided by the user or predefined to describe changes in the text modality that are applied to the visual modality, $M_t$ is derived entirely from the text modality and provides more specific and explicit modifications. At the same time, $\mathcal{A}_v^i$ produces $M_v$, which denotes a set indicating the presence or absence of visual attributes. The example of $M_v$ and $M_t$ can be found in Figure~\ref{fig:framework}. The outputs $M_t$ and $M_v$ will later be used by the question agent, which will be discussed in detail in a subsequent section.

\subsection{Coarse Filtering}
In Lines~5–9, each candidate image $I \in \mathcal{I}$ is evaluated by the text similarity-based scoring agent $\mathcal{A}_t^s$ and the vision similarity-based scoring agent $\mathcal{A}_v^s$. To improve robustness, the scoring process integrates three levels of cross-modality, providing multiperspective information across different stages and producing more reliable results through hybrid cross-modal similarity.

First, the scoring of both agents relies on $C_t$ and $C_v$ produced by $\mathcal{A}_t^i$ and $\mathcal{A}_v^i$, which together approximate the ideal target image $I_i$. Although both $C_t$ and $C_v$ are textual representations, they are derived from different modalities: the text modality and the vision modality, respectively. Second, each candidate image indexed by $a \in \{1, \ldots, N\}$ is also evaluated with respect to its caption $C_a$ and its visual content $I_a$ by $\mathcal{A}_t^s$ and $\mathcal{A}_v^s$, respectively. Third, each scoring agent produces two cross-modal scores. In Line~6, the scores $s_t^t$ and $s_t^v$ are generated by the text similarity-based scoring agent $\mathcal{A}_t^s$. Specifically, $s_t^t$ measures the similarity between $C_t$ and $C_a$, while $s_t^v$ measures the similarity between $C_v$ and $C_a$. Since $C_t$ originates from the text modality and $C_v$ implicitly reflects visual content, these two scores capture cross-modal signals that combine implicitly coupled and explicitly decoupled multimodality. Here, explicitly decoupled multimodality refers to processing that occurs entirely within the text modality, whereas implicit coupling indicates that visual information is embedded within the textual representation. The same procedure is applied by the vision similarity-based scoring agent $\mathcal{A}_v^s$. In Line~7, the scores $s_v^t$ and $s_v^v$ measure the similarity between $I_a$ and $C_t$, and between $I_a$ and $C_v$, respectively. Here, the pair $(I_a, C_v)$ belongs to the visual modality, whereas the pair $(I_a, C_t)$ combines the vision and text modalities. These procedures are collectively referred to as cross-modal scoring. Each score is generated by its corresponding agent, which encodes the inputs using an MLLM and computes the cosine similarity between the paired representations.

In Lines~8 and 9, for each candidate $I$, the scores from the text and vision modalities across agents are summed to obtain $s^t$ and $s^v$, respectively. Rather than aggregating the outputs of a single agent, the scores are aligned within each modality to ensure consistency in cross-modal scoring. In Line~10, a reciprocal rank fusion function is applied to transform the similarity score vectors $S^t = (s^t_1, \dots, s^t_N)$ and $S^v = (s^v_1, \dots, s^v_N)$ into rank values, which are then summed across the text and vision modalities. The ranking function is defined as:

\begin{equation}
    \operatorname{RRF}\left(a\right)=\frac{1}{z+\operatorname{rank}\left(s_a^t\right)}+\frac{1}{z+\operatorname{rank}\left(s_a^v\right)},
\end{equation}

\noindent where $\operatorname{rank}(s_a^t)$ and $\operatorname{rank}(s_a^v)$ denote the rank positions of the text-based score $s_a^t$ and the vision-based score $s_a^v$ among all candidates, respectively, and $z$ is a smoothing constant. This formulation ensures that higher-ranked candidates in either modality contribute more to the final score, while still incorporating signals from both modalities. Finally, the top-$k^\prime$ candidates $\mathcal{I}^{k^\prime}$ are selected and passed to the next stage for fine filtering, balancing retrieval accuracy with computational cost. If $k^\prime = N$, then all candidates are selected and proceed to the next step.

Notably, the process of producing and aggregating multiple scores is also inspired by human cognitive mechanisms. When humans search for target images, cross-modal information is interwoven in the mind, collectively forming a unified body of evidence that supports decision-making.

\begin{table*}[]
\caption{Performance comparison on FashionIQ validation set. The best results are in bold, and the second best are underlined.}
\vskip -2ex
\label{tab:fashioniq}
\resizebox{0.95\textwidth}{!}{
\begin{tabular}{c|llc|rr|rr|rr|rr}
\bottomrule  
\multicolumn{1}{c|}{\multirow{2}{*}{\textbf{Backbone}}} & \multicolumn{1}{c}{\multirow{2}{*}{\textbf{Method}}} & \multicolumn{1}{c}{\multirow{2}{*}{\textbf{Venue}}} & \multicolumn{1}{c|}{\multirow{2}{*}{\textbf{Training-free}}} & \multicolumn{2}{c|}{\textbf{Shirt}} & \multicolumn{2}{c|}{\textbf{Dress}} & \multicolumn{2}{c|}{\textbf{Toptee}} & \multicolumn{2}{c}{\textbf{\textit{Avg.}}} \\
\cline{5-6}  \cline{7-8} \cline{9-10}  \cline{11-12} 
\multicolumn{1}{c|}{} & \multicolumn{1}{c}{} & \multicolumn{1}{c}{}  & \multicolumn{1}{c|}{} & \multicolumn{1}{c}{R@10} & \multicolumn{1}{c|}{R@50} & \multicolumn{1}{c}{R@10} & \multicolumn{1}{c|}{R@50} & \multicolumn{1}{c}{R@10} & \multicolumn{1}{c|}{R@50} & \multicolumn{1}{c}{R@10} & \multicolumn{1}{c}{R@50} \\
\hline
\multirow{9}{*}{\rotatebox{90}{CLIP-ViT-B/32}} 
 & PALAVRA~\cite{Cohen2022ThisIM} & \textit{ECCV 2022} & \usym{1F5F6} & 21.49 & 37.05 & 17.25 & 35.94 & 20.55 & 38.76 & 19.76 & 37.25 \\
 & SEARLE~\cite{Baldrati2023ZeroShotCI} & \textit{ICCV 2023}  & \usym{1F5F6} & 24.44 & 41.61 & 18.54 & 39.51 & 25.70 & 46.46 & 22.89 & 42.53 \\
 & SEARLE-OTI~\cite{Baldrati2023ZeroShotCI} & \textit{ICCV 2023}  & \usym{1F5F6} & 25.37 & 41.32 & 17.85 & 39.91 & 24.12 & 45.79 & 22.45 & 42.34 \\
 & iSEARLE~\cite{iSEARLE} & \textit{TPAMI 2025} & \usym{1F5F6}  & 25.81 & 43.52 & 20.92 & 42.19 & 26.47 & 48.70 & 24.40 & 44.80 \\
 & iSEARLE-OTI~\cite{iSEARLE} & \textit{TPAMI 2025} & \usym{1F5F6}  & 27.09 & 43.42 & 21.27 & 42.19 & 26.82 & 48.75 & 25.06 & 44.79 \\
 & CIReVL~\cite{CIReVL} & \textit{ICLR 2024} & \usym{2714} & \underline{28.36} & \underline{47.84} & \underline{25.29} & \underline{46.36} & \underline{31.21} & \underline{53.85} & \underline{28.29} & \underline{49.35} \\
 & LDRE~\cite{yang2024ldre} & \textit{SIGIR 2024} & \usym{2714} & 27.38 & 46.27 & 19.97 & 41.84 & 27.07 & 48.78 & 24.81 & 45.63 \\
 & ImageScope~\citep{luo2025imagescope} & \textit{WWW 2025} & \usym{2714} & 24.29 & 37.49 & 18.00 & 35.20 & 24.99 & 41.41 & 22.42 & 38.03 \\
  \rowcolor{lightblue}
  & \textbf{$\text{X}^\text{R}$(Ours)} & \textbf{\textit{Proposed}} & \usym{2714} &  \textbf{36.06} & \textbf{54.66} & \textbf{30.94} & \textbf{52.06} & \textbf{42.99} & \textbf{64.56} & \textbf{36.66} & \textbf{57.10} \\
\hline
\multirow{11}{*}{\rotatebox{90}{CLIP-ViT-L/14}} 
 & Pic2Word~\cite{saito2023pic2word} & \textit{CVPR 2023} & \usym{1F5F6} & 26.20 & 43.60 & 20.00 & 40.20 & 27.90 & 47.40 & 24.70 & 43.73 \\
 & SEARLE~\cite{Baldrati2023ZeroShotCI} & \textit{ICCV 2023}  & \usym{1F5F6} & 26.89 & 45.58 & 20.48 & 43.13 & 29.32 & 49.97 & 25.56 & 46.23 \\
 & SEARLE-OTI~\cite{Baldrati2023ZeroShotCI} & \textit{ICCV 2023}  & \usym{1F5F6} & 30.37 & 47.49 & 21.57 & 44.47 & 30.90 & 51.76 & 27.61 & 47.91 \\
 & iSEARLE~\cite{iSEARLE} & \textit{TPAMI 2025} & \usym{1F5F6}  & 28.75 & 47.84 & 22.51 & 46.36 & 31.31 & 52.68 & 27.52 & 48.96 \\
 & iSEARLE-OTI~\cite{iSEARLE}& \textit{TPAMI 2025} & \usym{1F5F6} & \underline{31.80} & 50.20 & 24.19 & 45.12 & 31.72 & 53.29 & 29.24 & 49.54 \\
  & LinCIR~\cite{LinCIR} & \textit{CVPR 2024} & \usym{1F5F6} & 29.10 & 46.81 & 20.92 & 42.44 & 28.81 & 50.18 & 26.28 & 46.48 \\
 & FTI4CIR~\cite{Lin2024FinegrainedTI} &  \textit{SIGIR 2024} & \usym{1F5F6} & {31.35} & 50.59 & 24.49 & \underline{47.84} & {32.43} & \underline{54.21} & \underline{29.42} & \underline{50.88} \\
 & CIReVL~\cite{CIReVL} & \textit{ICLR 2024} & \usym{2714} & 29.49 & 47.40 & \underline{24.79} & 44.76 & 31.36 & 53.65 & 28.55 & 48.57 \\
 & LDRE~\cite{yang2024ldre} & \textit{SIGIR 2024} & \usym{2714} & 31.04 & \underline{51.22} & 22.93 & {46.76} & 31.57 & 53.64 & 28.51 & 50.54 \\
 & ImageScope~\citep{luo2025imagescope} & \textit{WWW 2025} & \usym{2714} & 27.82 & 41.76 & 20.18 & 37.48 & 28.61 & 44.42 & 25.54 & 41.22 \\
  \rowcolor{lightblue}
  & \textbf{$\text{X}^\text{R}$(Ours)} & \textbf{\textit{Proposed}} & \usym{2714} & \textbf{38.91} & \textbf{56.82} & \textbf{28.71} & \textbf{52.50} & \textbf{43.91} & \textbf{62.57} & \textbf{37.18} & \textbf{57.30} \\
\toprule
\end{tabular}}
\vspace{-2mm}
\end{table*}

\subsection{Fine Filtering}
The previously discussed efforts for CIR primarily focus on approximating $I_i$ through similarity-based scoring. However, in real-world scenarios, ensuring retrieval accuracy often requires factual verification with semanic reasoning. 

To address this, we introduce the question agent $\mathcal{A}^q$ (Line~11). This agent operates on three types of instructions: $M_t$, which represents modifications derived from $\mathcal{A}_t^i$; $M_v$, which represents visual attribute indicators; and $T_m$, which encodes atomic user-specified instructions. The agent $\mathcal{A}^q$ generates a set of verification questions $Q$ with corresponding answers $A$ by transforming each atomic instruction in $T_m$ into a declarative statement, while using $M_t$ as supporting context and $M_v$ as factual grounding. When $T_m$ alone is insufficient to define a clear question, $M_t$ provides additional context. All questions are formulated in a True/False format, with examples provided in Figure~\ref{fig:framework}.

In Lines~13 and~14, $\mathcal{A}_t^q$ and $\mathcal{A}_v^q$ are two question-based scoring agents that use $Q$ to evaluate each candidate image $a \in \{1, \dots, k^\prime\}$ based on different modality information, namely $C_a$ and $I_a$. We consider a fact to be true if the candidate is able to pass the verification check under both modalities. If the answer to a question is correct, the agent assigns a score of $+1$; otherwise, the score is $0$. The results are denoted as $s_q^t$ and $s_q^v$. We consider this design to be effective because it provides discrete, verifiable signals that emphasize factual consistency across modalities.

\begin{table*}[]
\caption{Ablation studies on CLIP-ViT-B/32 with InternVL3-8B.}
\vskip -2ex
\resizebox{\linewidth}{!}{
\begin{tabular}{cccc cc ccc ccc ccc}
\toprule
\multicolumn{2}{c}{\textbf{Similarity-based}} & 
\multicolumn{2}{c}{\textbf{Question-based}} & 
\multicolumn{2}{c}{\textbf{FashionIQ}} & 
\multicolumn{3}{c}{\textbf{CIRCO}} & 
\multicolumn{3}{c}{\textbf{CIRR}} & 
\multicolumn{3}{c}{\textbf{CIRR$_{subset}$}} \\
\cmidrule(lr){1-2} \cmidrule(lr){3-4} \cmidrule(lr){5-6} \cmidrule(lr){7-9} \cmidrule(lr){10-12} \cmidrule(lr){13-15}
\textbf{Textual} & \textbf{Visual} & \textbf{Textual} & \textbf{Visual} & R@10 & R@50  & mAP@5 & mAP@10 & mAP@25 & R@1 & R@5 & R@10 & R@1 & R@2 & R@3  \\
\midrule
\usym{1F5F6} & \usym{1F5F6} & \usym{1F5F6} & \usym{1F5F6} & 14.78 & 29.60 & 2.65 & 3.25 & 4.14 & 11.71 & 35.06 & 48.94 & 32.77 & 56.89 & 74.96 \\
\usym{2714} & \usym{1F5F6} & \usym{1F5F6} & \usym{1F5F6} & 19.36 & 37.65 & 11.98 & 13.40 & 14.11 &  18.12 & 51.16 & 65.11 & 59.76 & 79.88 & 90.00\\
\usym{1F5F6} & \usym{2714} & \usym{1F5F6} & \usym{1F5F6} & 32.48 & 54.55 & 15.18 & 16.02 & 17.54 & 27.02 & 61.04 & 74.05 & 64.53 & 83.18 & 91.93 \\
\usym{2714} & \usym{2714} & \usym{1F5F6} & \usym{1F5F6} & 32.84 & 55.37 & 16.73 & 17.69 & 19.29 & 27.33 & 63.57 & 76.36 & 66.39 & 84.00 & 92.89\\
\usym{2714} & \usym{1F5F6} & \usym{2714} & \usym{1F5F6} & 23.93 & 41.72 & 17.22 & 17.7 & 19.05 & 30.53 & 58.96 & 69.78 & 68.12 & 83.88 & 91.64\\
\usym{1F5F6} & \usym{2714} & \usym{1F5F6} & \usym{2714} &  36.01 & 56.57 & 24.12 & 24.84 & 26.53 & 40.24 & 71.57 & 81.75 & 75.42 & 89.68 & 93.77\\
\usym{2714}  & \usym{2714} & \usym{2714} & \usym{1F5F6} & 34.78 & 55.63 & 24.87 & 25.51 & 27.34 & 36.60 & 65.69 & 76.72 & 72.72 & 86.41 & 93.64\\
\usym{2714}  & \usym{2714} & \usym{1F5F6} & \usym{2714} &  \underline{36.62} & \underline{56.84} & \underline{26.21} & \underline{27.01} & \underline{28.87} & \underline{41.34} & \underline{73.06} & \underline{82.43} & \underline{76.45} & \underline{89.95} & \underline{95.13} \\
 \rowcolor{lightblue}
\usym{2714}  & \usym{2714}  & \usym{2714} & \usym{2714} &\textbf{36.66} & \textbf{57.10} & \textbf{27.51} & \textbf{28.33} & \textbf{30.28}  & \textbf{43.06} & \textbf{73.86} & \textbf{83.15} & \textbf{77.54} & \textbf{90.27} & \textbf{95.21}\\
\bottomrule
\end{tabular}
}
\label{tab:ablation}
\vspace{-2mm}
\end{table*}

In Lines~15 and~16, we define the cross-modal re-ranking procedure for the top-$k^\prime$ previously selected candidates. For notational simplicity, we omit the index $k^\prime$ here. First, the question-based scores for each candidate are summed as $S_q^t + S_q^v$, where $s_q^t \in S_q^t$, $s_q^v \in S_q^v$, and $|S_q^t| = |S_q^v| = k^\prime$. This sum is then multiplied by a normalized weighted combination of the similarity-based scores $S^t$ and $S^v$ from Lines~8 and~9, with weights $\lambda$ and $1-\lambda$, respectively. It is worth noting that only the similarity-based scores $S^t$ and $S^v$ of the top-$k^\prime$ candidates are used at this stage. The result, denoted $S^{k^\prime}$, represents the refined scores used for re-ranking in Line~16. Finally, the re-ranked set $\mathcal{I}^*$ is returned as the final output.

This design is motivated by the complementary strengths of the two scoring mechanisms. Similarity-based scores ($S^t$ and $S^v$) capture soft alignment between the candidate images and the multimodal query, but they may overlook fine-grained factual details. In contrast, question-based scores ($S_q^t$ and $S_q^v$) enforce explicit verification of atomic modifications and provide binary, interpretable signals. By combining the two, the re-ranking step integrates the broad cross-modal coverage of similarity-based scoring with the precision of question-based verification, thereby improving robustness and ensuring that the final retrieved ordered set $\mathcal{I}^*$ more faithfully reflects the intended modifications through semantic reasoning.

In fact, the use of $\text{X}^\text{R}$ is flexible. So far, we have discussed how similarity-based scoring and question-based scoring collaborate through ranking (selection) and re-ranking (re-selection). Moreover, when $k^\prime = k$, the two scoring processes act as ranking–selection and re-ranking. When $k^\prime = N$, the two processes operate jointly, performing ranking and selection directly. It is important to note that these three configurations correspond to increasing computational cost.

In summary, we propose $\text{X}^\text{R}$, a training-free cross-modal multi-agent framework for CIR. The framework highlights the benefits of agentic AI, including minimal human intervention and autonomous collaboration among multiple agents that mimic human cognitive processes. By enabling retrieval that is both robust and adaptive across modalities, $\text{X}^\text{R}$ points toward promising directions for large-scale information access scenarios that are becoming increasingly central in web-driven environments.
\section{Experiments}
\label{sec:experiment}

\subsection{Experiment Setup}
\begin{table}[]
\caption{Benchmark details.}
\vspace{-3ex}
\resizebox{\columnwidth}{!}{
\begin{tabular}{lccrr}
\toprule
\multicolumn{1}{c}{\textbf{Dataset}} & \textbf{Split} & \textbf{Type} & \multicolumn{1}{c}{\textbf{\# Queries}} & \multicolumn{1}{c}{\textbf{\# Images}} \\
\midrule
CIRR~\cite{Liu2021ImageRO} & Test & CIR & 4,148 & 2,316 \\
CIRCO~\cite{Baldrati2023ZeroShotCI} & Test & CIR & 800 & 123,403 \\
FashionIQ-Shirt~\cite{wu2021fashion} & Val. & CIR & 2,038 & 6,346 \\
FashionIQ-Dress~\cite{wu2021fashion} & Val. & CIR & 2,017 & 3,817 \\
FashionIQ-Toptee~\cite{wu2021fashion} & Val. & CIR & 1,961 & 5,373 \\
\bottomrule
\end{tabular}
}
\label{table:benchmark_details}
\vskip -2ex
\end{table}
\noindent \textbf{Benchmark.}
We evaluate $\text{X}^\text{R}$ on three representative CIR benchmarks (Table~\ref{table:benchmark_details}):  
CIRR~\citep{Liu2021ImageRO}, the first natural-image CIR dataset with subset retrieval for fine-grained candidate discrimination;  
CIRCO~\citep{Baldrati2023ZeroShotCI}, a large-scale benchmark with multiple ground truths to reduce false negatives;  
and FashionIQ~\citep{wu2021fashion}, a fashion-domain dataset with three categories (dress, shirt, toptee).

\noindent \textbf{Metrics.}
Following the original protocols, we use Recall@k (R@k) for CIRR and FashionIQ to capture retrieval accuracy, and mean average precision (mAP@k) for CIRCO to account for multiple valid ground truths.

\noindent \textbf{Baselines.}
We compare $\text{X}^\text{R}$ against nine representative CIR baselines. 
Since $\text{X}^\text{R}$ is training-free, we primarily focus on zero-shot methods for fair comparison, while also reporting strong training-based models for completeness:

\begin{itemize}
\setlength\itemsep{0pt}
\setlength\parsep{0pt}
\setlength\topsep{0pt}
    \item \textbf{Training-based:} PALAVRA~\citep{Cohen2022ThisIM}, Pic2Word~\citep{saito2023pic2word}, SEARLE~\citep{Baldrati2023ZeroShotCI}, iSEARLE~\citep{iSEARLE}, LinCIR~\citep{LinCIR}, and FTI4CIR~\citep{Lin2024FinegrainedTI}.
    \item \textbf{Training-free:} CIReVL~\citep{CIReVL}, LDRE~\citep{yang2024ldre}, and ImageScope~\citep{luo2025imagescope}.
\end{itemize}

\noindent \textbf{Implementation Details.} 
We use CLIP-ViT-L/14 and CLIP-ViT-B/32~\citep{Jia2021ScalingUV} as dual-encoder backbones for similarity agents, and InternVL3-8B~\citep{internvl3} for imagination and question-based verification.
We set the temperature $= 0$ and top-$p= 1$ for deterministic outputs.
The fusion weight is $\lambda=0.15$ to balance text–image similarity.
We set $k^\prime=100$ candidates for fine filtering to allow question-based scoring to take effect. 
If $k^\prime$ were set equal to $k$, as in Recall@k, the benefit of re-ranking would be masked since Recall@k is order-insensitive. 
In contrast, fine filtering both re-ranks and prunes candidates, making $k^\prime=100$ a balanced choice.
All experiments are conducted on a single NVIDIA H800-80G GPU with FP16 precision. 

\begin{figure*}[t]
  \centering
  \includegraphics[width=\textwidth]{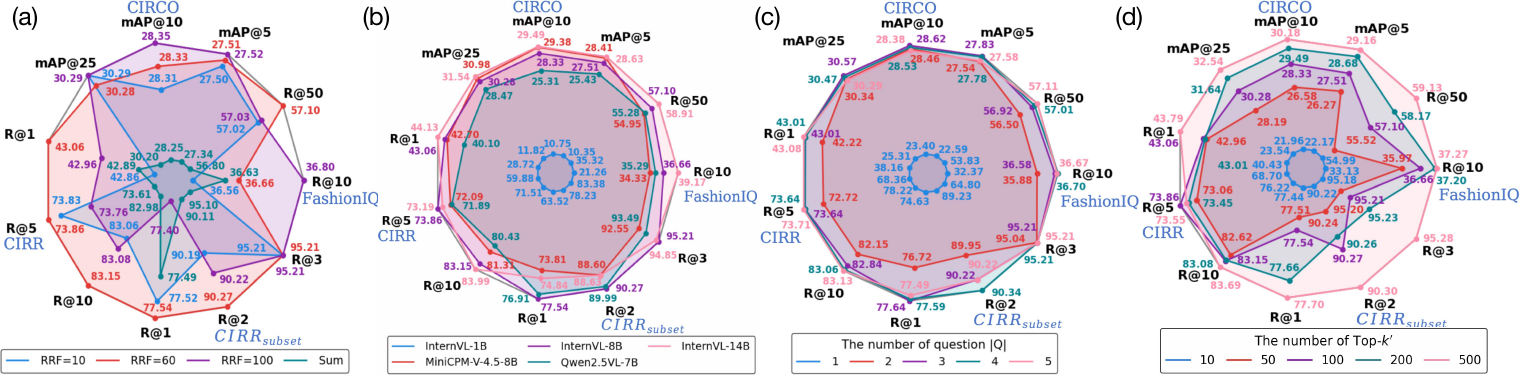}
  \vskip -2ex
  \caption{Parameter analysis of $\text{X}^\text{R}$. 
(a) Effect of RRF with different $z$ values. 
(b) Comparison across multimodal backbones. 
(c) Impact of the number of verification questions. 
(d) Influence of candidate pool size $k^\prime$.}
\label{fig:parameter}
\end{figure*}

\begin{figure}[t]
  \centering
  \includegraphics[width=0.75\linewidth]{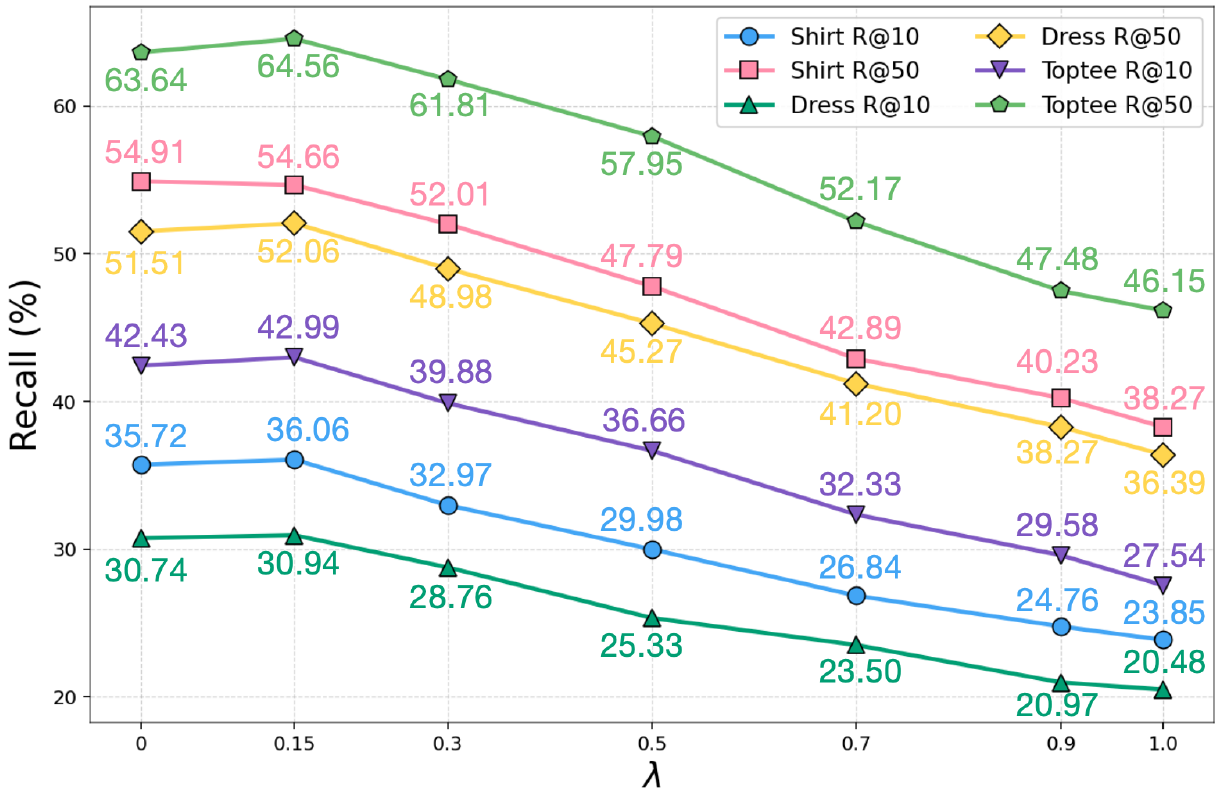}
    \vskip -2ex
  \caption{Effect of $\lambda$ on text–image fusion: best at $\lambda{=}0.15$; extremes degrade by losing cross-modal cues.}
  \label{fig:lambda}
\end{figure}

\subsection{Main Result.}
Table~\ref{tab:cirr_circo} and Table~\ref{tab:fashioniq} illustrate that $\text{X}^\text{R}$ consistently outperforms both training-free and training-based CIR methods.
On FashionIQ, $\text{X}^\text{R}$ achieves consistent gains across all three categories. 
With CLIP-ViT-B/32, it reaches 36.66\% R@10 and 57.10\% R@50 on average, surpassing CIReVL by over 8 points in R@10. 
These gains hold across shirts, dresses, and tops, indicating that the method generalizes across diverse attribute-level edits rather than overfitting to a single category. On CIRCO, which introduces large distractor sets and multiple ground truths, $\text{X}^\text{R}$ attains 30.95\% mAP@50, over 7 points higher than the best baseline. 
This shows that multi-agent reasoning maintains robustness in noisy, large-scale retrieval where static similarity models often collapse. 
On CIRR, $\text{X}^\text{R}$ achieves 83.15\% R@10 and 95.21\% R@3 in the fine-grained subset retrieval task ($\text{CIRR}_{\text {subset }}$), surpassing training-free and training-based baselines. 
These results show that imagination and verification act as complementary safeguards against error propagation, ensuring faithful alignment to user intent in fine-grained scenarios. 
Overall, $\text{X}^\text{R}$ demonstrates consistent advantages across domain-specific (FashionIQ), distractor-heavy (CIRCO), and fine-grained (CIRR) benchmarks, pointing to strong cross-benchmark generalization.

\subsection{Ablation Studies.}
\noindent \textbf{Ablation on key modules.}
Table~\ref{tab:ablation} presents results from progressively enabling different modules. On FashionIQ, the visual similarity agent alone lifts R@10 from 14.78\% to 32.48\%. Compared with the textual agent, the two together show clear complementarity: the visual branch captures appearance-level cues but risks semantic drift, while the textual branch enforces semantic alignment but may miss subtle details. Combining both further yields substantial gains, with CIRCO mAP@25 rising from 4.14\% to 19.29\%, confirming that cross-modal similarity is more reliable than unimodal signals. Adding a textual question-based agent then further boosts CIRR R@10 to 76.72\%, showing that factual checks reduce false positives. Finally, with both textual and visual question-based agents, CIRR$_{subset}$ R@3 reaches 95.21\%, demonstrating that explicit verification enforces modification faithfulness. 
In summary, each module contributes individually, but their integration ultimately delivers the strongest performance: similarity agents provide broad alignment, and question-based agents enforce correctness, together validating the multi-agent cross-modal reasoning of $\text{X}^\text{R}$.

\subsection{Discussion.}

\noindent \textbf{RRF $z$ vs.\ summation.}  
Figure~\ref{fig:parameter}(a) compares reciprocal rank fusion (RRF) with varying $z$ against direct score summation. Direct score summation performs worst, showing that naive averaging fails to normalize heterogeneous modalities. 
RRF instead focuses on ranks, which makes aggregation more robust to noisy candidates. 
At $z=60$, it strikes the best balance, raising CIRCO mAP@25 to 30.28\% and CIRR R@10 to 83.15\%.

\noindent \textbf{Generality of MLLMs.}  
Figure~\ref{fig:parameter}(b) compares different multimodal backbones. 
Medium-scale models such as InternVL3-8B and Qwen2.5VL-7B achieve the best trade-off, combining strong grounding ability with efficiency (e.g., 57.10\% R@50 on FashionIQ and 95.21\% R@3 on CIRR$_{subset}$). Smaller backbones lack grounding ability, while extremely large ones bring only marginal gains at a much higher cost. This suggests that $\text{X}^\text{R}$ benefits most from medium-scale MLLMs, which balance expressiveness and efficiency and indicate that the framework scales smoothly across backbones.

\noindent \textbf{Number of questions.}  
Figure~\ref{fig:parameter}(c) shows that a single verification question is insufficient, while three yield consistent gains across benchmarks (e.g., CIRR R@10 = 82.84\%). Using more than three slightly reduces performance, as redundant checks add overhead without new information. 
These results indicate that a small, diverse set of questions suffices for robust factual alignment, consistent with the principle that each agent contributes distinct value.

\noindent \textbf{Top-$k^\prime$ analysis.}  
Figure~\ref{fig:parameter}(d) illustrates the impact of the coarse filtering pool size. Small $k^\prime$ limits recall, while larger pools improve coverage by ensuring candidate diversity. Gains plateau beyond $k^\prime=100$, with only marginal improvements up to $k^\prime=500$. This indicates that moderately large pools provide the best efficiency–effectiveness balance.

\noindent \textbf{Effect of $\lambda$.}  
Figure~\ref{fig:lambda} examines the balance between textual and visual similarity signals. Relying solely on either modality ($\lambda=0$ or $\lambda=1$) significantly degrades performance, as it ignores complementary cues. 
The best results occur at $\lambda=0.15$, suggesting that cross-modal fusion is most effective when neither modality dominates, validating the principle of balanced agent collaboration.

\begin{figure}[!t]
  \centering
  \begin{subfigure}{0.49\linewidth}
    \centering
    \includegraphics[width=\linewidth]{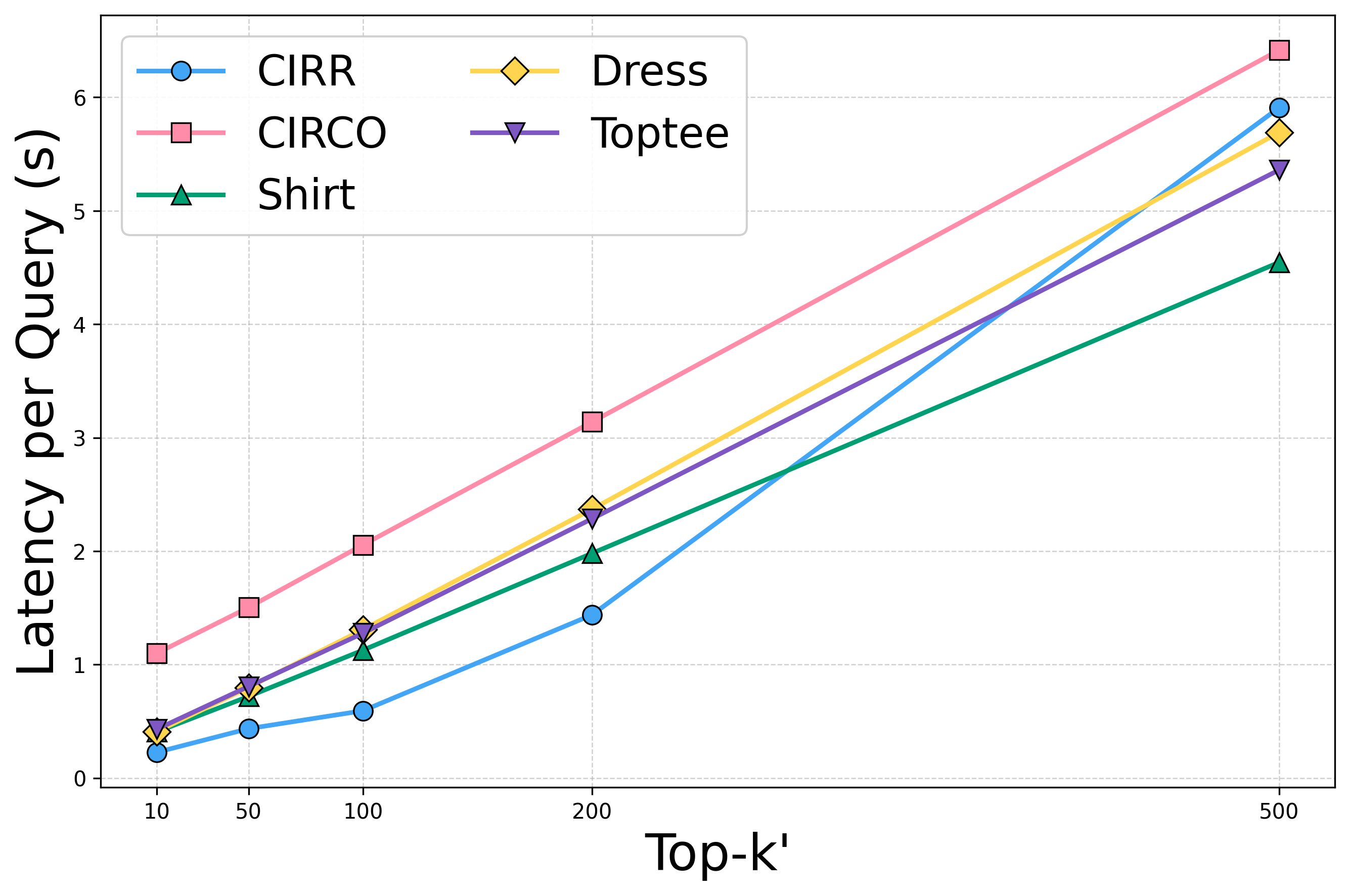}
    \caption{Average latency per query.}
    \label{fig:latency_avg}
  \end{subfigure}
  \hfill
  \begin{subfigure}{0.49\linewidth}
    \centering
    \includegraphics[width=\linewidth]{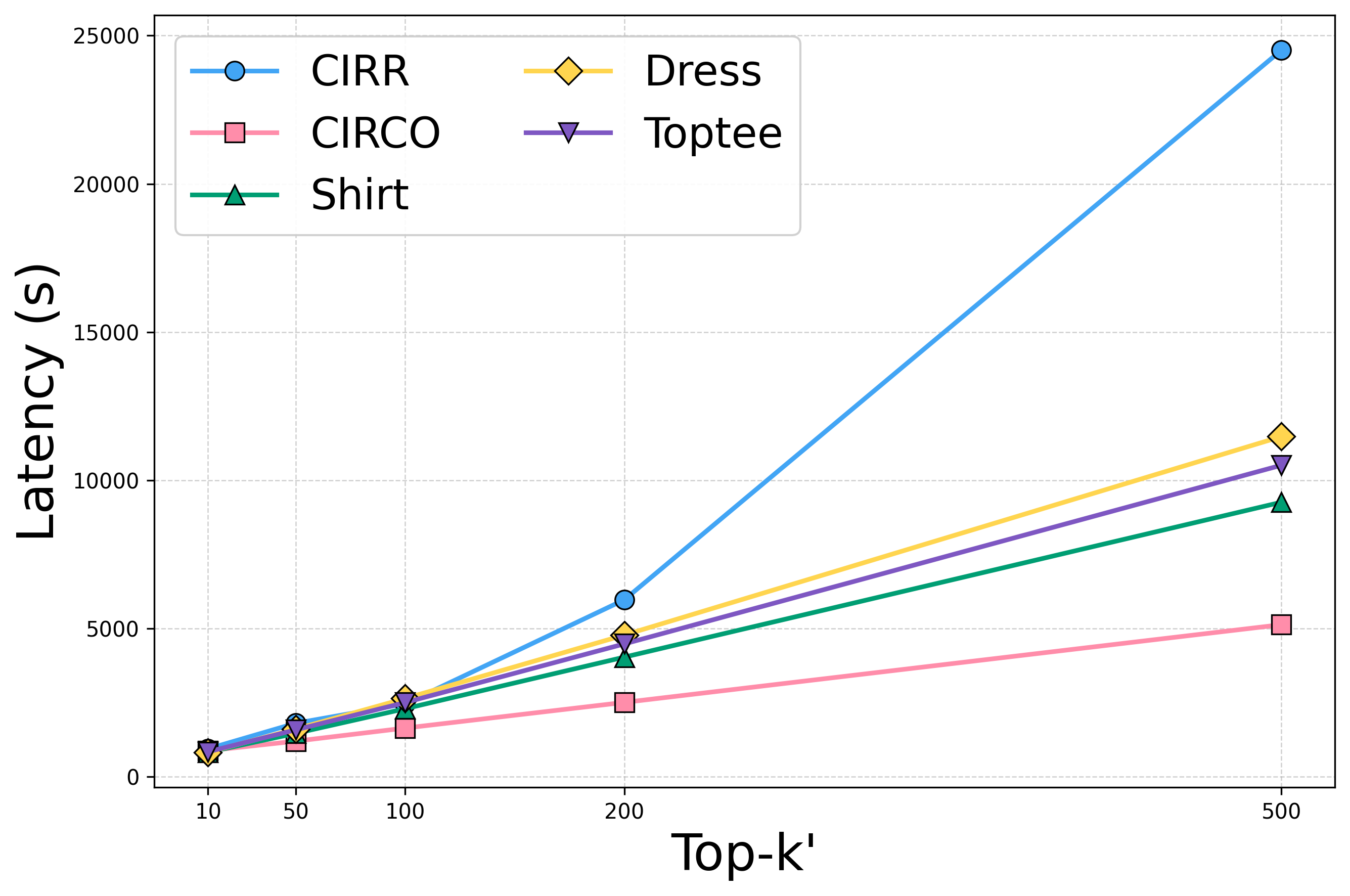}
    \caption{Total latency.}
    \label{fig:latency_total}
  \end{subfigure}
  \vskip -2ex
  \caption{Latency of $\text{X}^\text{R}$ under different top-$k^\prime$: larger pools increase cost nearly linearly, but $k^\prime \approx 100$ balances coverage and overhead.}
  \label{fig:latency_analysis}
\end{figure}

\noindent \textbf{Latency analysis.}  
Figure~\ref{fig:latency_analysis} shows average and total latency under different candidate pool sizes ($k'$). As expected, larger $k'$ increases cost as more candidates enter fine filtering. Per-query latency grows nearly linearly, with CIRCO highest due to its many distractors. Total latency is dominated by CIRR because of its dataset scale, while FashionIQ remains relatively lightweight. Overall, a moderately large $k'$ (around 100) offers the best trade-off, balancing diversity for robust retrieval against computational overhead. 

These results highlight that the strength of $\text{X}^\text{R}$ comes from orchestrating multiple agents rather than relying on any single component. 
By uniting semantic alignment with factual verification, cross-modal reasoning refines retrieval and overcomes the inherent limits of unimodal pipelines. 
More broadly, this shows that cross-modal multi-agent reasoning is not only effective for CIR but establishes a general paradigm for multimodal retrieval and reasoning.

\section{Conclusion}
We presented $\text{X}^\text{R}$, a training-free cross-modal multi-agent framework for composed image retrieval. 
Unlike unimodal pipelines, $\text{X}^\text{R}$ integrates imagination, coarse filtering, and fine filtering through similarity- and question-based agents, progressively refining results via semantic alignment and factual verification. 
Experiments on FashionIQ, CIRCO, and CIRR show consistent improvements over both training-free and training-based baselines, particularly in fine-grained and distractor-rich scenarios. 
Ablation analyses confirm that while each agent contributes independently, their coordination yields more stable and accurate retrieval. 
Overall, our findings underscore that cross-modal reasoning is not only advantageous but often essential for aligning retrieval with user intent. 
Looking ahead, we envision $\text{X}^\text{R}$ as a foundation for retrieval-augmented reasoning, where agentic systems actively interpret, verify, and adapt across modalities to achieve reliable and human-aligned intelligence.

\newpage
\bibliographystyle{ACM-Reference-Format}
\bibliography{main}

\appendix
\newpage
\begin{center}
  \LARGE
  \textbf{Appendix} 
\end{center}

\renewcommand{\thetheorem}{A.\arabic{theorem}}
\renewcommand{\thefigure}{A.\arabic{figure}}
\renewcommand{\thetable}{A.\arabic{table}}
\renewcommand{\theequation}{A.\arabic{equation}}
\setcounter{figure}{0}
\setcounter{theorem}{0}
\setcounter{table}{0}
\setcounter{equation}{0}

\begin{table*}[!ht]
\caption{Ablation studies on CLIP-ViT-B/32 with InternVL3-8B on FashionIQ.}
\vskip -2ex
\resizebox{0.85\linewidth}{!}{
\begin{tabular}{cccc cc cc cc cc}
\toprule
\multicolumn{2}{c}{\textbf{Similarity-based}} & 
\multicolumn{2}{c}{\textbf{Question-based}} & \multicolumn{2}{c}{\textbf{Shirt}} & \multicolumn{2}{c}{\textbf{Dress}} & \multicolumn{2}{c}{\textbf{Toptee}} & \multicolumn{2}{c}{\textbf{\textit{Avg.}}} 
 \\
\cmidrule(lr){1-2}  \cmidrule(lr){3-4} \cmidrule(lr){5-6}  \cmidrule(lr){7-8} \cmidrule(lr){9-10}  \cmidrule(lr){11-12} 
\textbf{Textual} & \textbf{Visual} & \textbf{Textual} & \textbf{Visual} & \multicolumn{1}{c}{R@10} & \multicolumn{1}{c}{R@50} & \multicolumn{1}{c}{R@10} & \multicolumn{1}{c}{R@50} & \multicolumn{1}{c}{R@10} & \multicolumn{1}{c}{R@50} & \multicolumn{1}{c}{R@10} & \multicolumn{1}{c}{R@50} \\
\hline
\usym{1F5F6} & \usym{1F5F6} & \usym{1F5F6} & \usym{1F5F6} & 13.44 & 26.25 & 13.83 & 30.88 & 17.08 & 31.67 & 14.78 & 29.60 \\
\usym{2714} & \usym{1F5F6} & \usym{1F5F6} & \usym{1F5F6} & 19.48 & 35.62 & 16.21 & 33.96 & 22.39 & 43.35 & 19.36 & 37.65 \\
\usym{1F5F6} & \usym{2714} & \usym{1F5F6} & \usym{1F5F6} &  32.92 & 53.19 & 26.13 & 49.18 & 38.40 & 61.30 & 32.48 & 54.55 \\
\usym{2714} & \usym{1F5F6} & \usym{2714} & \usym{1F5F6} & 23.87 & 39.63 & 19.84 & 37.60 & 28.06 & 47.93 & 23.93 & 41.72 \\
\usym{1F5F6} & \usym{2714} & \usym{1F5F6} & \usym{2714} &  35.62 & 54.41 & 29.90 & 51.66 & 42.51 & 63.64 & 36.01 & 56.57 \\
\usym{2714} & \usym{2714} & \usym{1F5F6} & \usym{1F5F6} & 33.45 & 53.54 & 27.37 & 50.69 & 37.68 & 61.86 & 32.84 & 55.37 \\
\usym{2714} & \usym{2714} & \usym{2714} & \usym{1F5F6} & 34.74 & 53.48 & 28.90 & 50.07 & 40.69 & 63.34 & 34.78 & 55.63 \\
\usym{2714} & \usym{2714} & \usym{1F5F6} & \usym{2714} & 36.16 & 54.17 & 31.04 & 51.96 & 42.65 & 64.41 & 36.62 & 56.84\\
 \rowcolor{lightblue}
\usym{2714}  & \usym{2714}  & \usym{2714} & \usym{2714}  
& 36.06 & 54.66 & 30.94 & 52.06 & 42.99 & 64.56 & 36.66 & 57.10 \\
\toprule
\end{tabular}
}
\end{table*}

\section{Detailed Experiment Results}
In this section, we provide the complete parameter analysis and ablation study results on the FashionIQ dataset. While the main paper reports the representative metrics (R@10 and R@50), here we include the full set of scores across different categories (shirts, dresses, and tops) and a wider range of evaluation metrics. These results further demonstrate the consistent advantages of our cross-modal framework over single-modality baselines.

\begin{table}[!ht]
\caption{Number of questions studies on CLIP-ViT-B/32 with InternVL3-8B on FashionIQ.}
\vskip -2ex
\resizebox{\linewidth}{!}{
\begin{tabular}{c cc cc cc cc}
\toprule
\multirow{2}{*}{\textbf{Question Num}} & \multicolumn{2}{c}{\textbf{Shirt}} & \multicolumn{2}{c}{\textbf{Dress}} & \multicolumn{2}{c}{\textbf{Toptee}} & \multicolumn{2}{c}{\textbf{\textit{Avg.}}} \\
\cmidrule(lr){2-3} \cmidrule(lr){4-5} \cmidrule(lr){6-7} \cmidrule(lr){8-9}
 & R@10 & R@50 & R@10 & R@50 & R@10 & R@50 & R@10 & R@50 \\
\midrule
1 & 32.14 & 51.67 & 26.38 & 49.33 & 38.60 & 60.53 & 32.37 & 53.83\\
2 & 34.79 & 54.51 & 30.00 & 51.41 & 42.84 & 63.59 & 35.88 & 56.50\\
3 & 35.82 & 54.66 & 31.18 & 52.01 & 42.73 & 64.10 & 36.58 & 56.92 \\
4 & 35.87 & 54.76 & 31.18 & 51.96 & 43.04 & 64.30 & 36.70 & 57.01\\
5 & 36.11 & 54.91 & 30.94 & 52.01 & 42.94 & 64.41 & 36.67 & 57.11\\
\bottomrule
\end{tabular}
}
\end{table}

\begin{table}[!ht]
\caption{Generality of MLLMs studies on CLIP-ViT-B/32 with InternVL3-8B on FashionIQ.}
\vskip -2ex
\resizebox{\linewidth}{!}{
\begin{tabular}{c cc cc cc cc}
\toprule
\multirow{2}{*}{\textbf{MLLMs}} & \multicolumn{2}{c}{\textbf{Shirt}} & \multicolumn{2}{c}{\textbf{Dress}} & \multicolumn{2}{c}{\textbf{Toptee}} & \multicolumn{2}{c}{\textbf{\textit{Avg.}}} \\
\cmidrule(lr){2-3} \cmidrule(lr){4-5} \cmidrule(lr){6-7} \cmidrule(lr){8-9}
 & R@10 & R@50 & R@10 & R@50 & R@10 & R@50 & R@10 & R@50 \\
\midrule
InternVL3-1B & 18.66 & 29.99 & 17.90 & 34.26 & 27.72 & 41.72 & 21.26 & 35.32 \\
InternVL3-8B & 36.06 & 54.66 & 30.94 & 52.06 & 42.99 & 64.56 & 36.66 & 57.10\\
InternVL3-14B & 38.86 & 56.82 & 33.47 &54.88 & 45.18 & 65.02 & 39.17 & 58.91\\
MiniCPM-V-4.5-8B & 33.52 & 54.40 & 30.51 & 50.33 & 41.84 & 61.12 & 35.29 & 55.28 \\
Qwen2.5VL-7B & 31.79 & 52.89 & 31.33 & 50.22 & 39.86 & 61.73 & 34.33 & 54.95 \\
\bottomrule
\end{tabular}
}
\end{table}

\begin{table}[!ht]
\caption{RRF $z$ vs.\ summation studies on CLIP-ViT-B/32 with InternVL3-8B on FashionIQ.}
\vskip -2ex
\resizebox{\linewidth}{!}{
\begin{tabular}{c cc cc cc cc}
\toprule
\multirow{2}{*}{\textbf{Method}} & \multicolumn{2}{c}{\textbf{Shirt}} & \multicolumn{2}{c}{\textbf{Dress}} & \multicolumn{2}{c}{\textbf{Toptee}} & \multicolumn{2}{c}{\textbf{\textit{Avg.}}} \\
\cmidrule(lr){2-3} \cmidrule(lr){4-5} \cmidrule(lr){6-7} \cmidrule(lr){8-9}
 & R@10 & R@50 & R@10 & R@50 & R@10 & R@50 & R@10 & R@50 \\
\midrule
RRF=10 & 35.97 & 55.10 & 30.94 & 52.06 & 42.78 & 63.90 & 36.56 & 57.02\\
RRF=60 & 36.06 & 54.66 & 30.94 & 52.06 & 42.99 & 64.56 & 36.66 & 57.10 \\
RRF=100 & 36.31 & 55.05 & 30.99 & 52.06 & 43.09 & 64.00 & 36.80 & 57.03\\
Sum & 35.91 & 54.35 & 30.99 & 52.06 & 42.99 & 64.00 & 36.63 & 56.80 \\
\bottomrule
\end{tabular}
}
\end{table}

\begin{table}[!ht]
\caption{Top-$k^\prime$ analysis on CLIP-ViT-B/32 with InternVL3-8B on FashionIQ.}
\vskip -2ex
\resizebox{\linewidth}{!}{
\begin{tabular}{c cc cc cc cc}
\toprule
\multirow{2}{*}{\textbf{Top-$k'$}} & \multicolumn{2}{c}{\textbf{Shirt}} & \multicolumn{2}{c}{\textbf{Dress}} & \multicolumn{2}{c}{\textbf{Toptee}} & \multicolumn{2}{c}{\textbf{\textit{Avg.}}} \\
\cmidrule(lr){2-3} \cmidrule(lr){4-5} \cmidrule(lr){6-7} \cmidrule(lr){8-9}
 & R@10 & R@50 & R@10 & R@50 & R@10 & R@50 & R@10 & R@50 \\
\midrule
10 & 33.37 & 53.73 & 26.87 & 48.98 & 39.16 & 62.26 & 33.13 & 54.99\\
50 & 35.77 & 53.29 & 29.77 & 51.30 & 42.38 & 61.96 & 35.97 & 55.52\\
100 & 36.06 & 54.66 & 30.94 & 52.06 & 42.99 & 64.56 & 36.66 & 57.10 \\
200 & 36.26 & 55.64 & 31.68 & 53.54 & 43.65 & 65.32 & 37.20 & 58.17 \\
500 & 36.31 & 56.87 & 31.63 & 54.88 & 43.86 & 65.63 & 37.27 & 59.13\\
\bottomrule
\end{tabular}
}
\end{table}

\begin{table}[!ht]
\caption{Latency analysis (in seconds)  across categories under different Top-$k'$.}
\vskip -3ex
\centering
\resizebox{\linewidth}{!}{
\begin{tabular}{c ccccc c}
\toprule
\textbf{Top-$k'$} & \textbf{CIRR} & \textbf{CIRCO} & \textbf{FashionIQ$_{Shirt}$} & \textbf{FashionIQ$_{Dress}$} & \textbf{FashionIQ$_{Toptee}$} & \textbf{Avg.}\\
\midrule
10   & 945   & 881   & 831  & 829  & 847   & 867 \\
50   & 1808  & 1203  & 1467 & 1611 & 1585  & 1535 \\
100  & 2464  & 1645  & 2298 & 2643 & 2507  & 2311 \\
200  & 5970  & 2511  & 4037 & 4781 & 4483  & 4356 \\
500  & 24506 & 5132  & 9262 & 11476& 10509 & 12177 \\
\bottomrule
\end{tabular}
}
\end{table}

\begin{table}[!ht]
\caption{Average latency per query (in seconds) across categories under different Top-$k'$.}
\vskip -3ex
\centering
\resizebox{\linewidth}{!}{
\begin{tabular}{c ccccc c}
\toprule
\textbf{Top-$k$} & \textbf{CIRR} & \textbf{CIRCO} & \textbf{FashionIQ$_{Shirt}$} & \textbf{FashionIQ$_{Dress}$} & \textbf{FashionIQ$_{Toptee}$} & \textbf{Avg.} \\
\midrule
10   & 0.228 & 1.101 & 0.408 & 0.411 & 0.432 & 0.516 \\
50   & 0.436 & 1.504 & 0.720 & 0.799 & 0.608 & 0.653 \\
100  & 0.594 & 2.056 & 1.128 & 1.310 & 1.278 & 1.273 \\
200  & 1.439 & 3.139 & 1.981 & 2.370 & 2.286 & 2.243 \\
500  & 5.908 & 6.415 & 4.545 & 5.690 & 5.359 & 5.983 \\
\bottomrule
\end{tabular}
}
\end{table}

\renewcommand{\thetheorem}{B.\arabic{theorem}}
\renewcommand{\thefigure}{B.\arabic{figure}}
\renewcommand{\thetable}{B.\arabic{table}}
\renewcommand{\theequation}{B.\arabic{equation}}
\setcounter{figure}{0}
\setcounter{theorem}{0}
\setcounter{table}{0}
\setcounter{equation}{0}
\section{Statistical Significance Study}
\begin{table}[!ht]
\centering
\caption{Statistical comparison on \textsc{FashionIQ} benchmark with average Recall@50. \textbf{StdDev} denotes standard deviation. Paired one-sided $t$-test and Wilcoxon signed-rank test $p$-values are reported ($\alpha = 5\%$).}
\vskip -2ex
\resizebox{0.85\linewidth}{!}{
\begin{tabular}{lcccc}
\toprule
\textbf{Method} & \textbf{Mean (\%)} & \textbf{StdDev} & \textbf{$t$-test $p$} & \textbf{Wilcoxon $p$}  \\
\midrule
Raw & 29.62 & 0.11 & $3.39 \times 10^{-26}$ & \multirow{4}{*}{$4.88 \times 10^{-4}$} \\
CIReVL & 49.06 & 0.23 & $3.94 \times 10^{-17}$ &  \\
ImageScope & 37.99 & 0.25 & $6.82 \times 10^{-21}$ &\\
\textbf{$\text{X}^\text{R}$ (Ours)} & \textbf{57.16} & \textbf{0.07} & -- &  \\
\bottomrule
\end{tabular}
}
\label{tab:significance}
\end{table}

We conduct 10 independent runs with different random seeds and compare our method against baseline models under identical settings. 
To assess statistical significance, we perform paired one-sided $t$-tests and Wilcoxon signed-rank tests. 
The null hypothesis ($\mathbf{H}_0$) states that $\text{X}^\text{R}$ performs equally or worse than the baseline, while the alternative hypothesis ($\mathbf{H}_1$) states that our method performs better. 
As shown in Table~\ref{tab:significance}, $\text{X}^\text{R}$ achieves the highest mean score (57.16 $\pm$ 0.07), compared to CIReVL (49.06 $\pm$ 0.23), ImageScope (37.99 $\pm$ 0.25), and the Raw baseline (29.62 $\pm$ 0.11). 
All comparisons yield $p$-values well below the threshold $\alpha = 0.05$ (e.g., $t$-test: $3.94 \times 10^{-17}$ against CIReVL), thereby allowing us to confidently reject $\mathbf{H}_0$ in favor of $\mathbf{H}_1$. 
These results confirm that $\text{X}^\text{R}$ consistently and significantly outperforms all baselines.

\section{Experimental Code}.
To promote transparency and ensure the reproducibility of our work, we will release all experimental code, datasets, and detailed tutorials necessary for replicating our experiments. Our goal is to make it straightforward for researchers and practitioners to reproduce our results, regardless of their technical background. Additionally, by providing comprehensive documentation and clear guidelines, we aim to facilitate the extension of our method to other models and architectures, enabling the broader research community to explore its potential applications and improvements. We believe that open and reproducible research is essential for advancing the field and fostering collaboration.

\section{Limitations and Future Work}
\label{sec:limitation}

While $\text{X}^\text{R}$ achieves strong results on CIR benchmarks, several limitations remain. The framework is currently tailored to image–text composition and has not yet been explored in settings involving richer modalities or temporal data. Its reliance on captions and verification questions generated by large models can also introduce subtle biases, which may affect consistency. Moreover, scaling to very large candidate pools requires further optimization of efficiency.  
Looking ahead, we envision cross-modal reasoning as the key avenue for progress. Extending $\text{X}^\text{R}$ beyond images and text to modalities such as video, audio, or interactive queries would open new opportunities for retrieval systems. Developing more lightweight and adaptive agents, together with diverse verification signals, could further enhance both robustness and scalability. These directions highlight the potential of cross-modal multi-agent reasoning as a general paradigm for future multimodal search.

\section{Ethical Considerations}

\textbf{Reliability and Transparency.} 
$\text{X}^\text{R}$ enhances retrieval reliability by coordinating imagination, similarity, and verification, reducing semantic drift and promoting more trustworthy multimodal systems. Its modular design decomposes decisions into interpretable stages, enabling auditing and analysis of system behavior.

\noindent \textbf{Responsible Data Use.} 
All experiments are conducted on publicly available datasets with proper licenses, ensuring compliance with ethical data standards.

\renewcommand{\thetheorem}{F.\arabic{theorem}}
\renewcommand{\thefigure}{F.\arabic{figure}}
\renewcommand{\thetable}{F.\arabic{table}}
\renewcommand{\theequation}{F.\arabic{equation}}
\setcounter{figure}{0}
\setcounter{theorem}{0}
\setcounter{table}{0}
\setcounter{equation}{0}

\section{Case Studies}

\begin{figure*}[!ht]
  \centering
  \includegraphics[width=\linewidth]{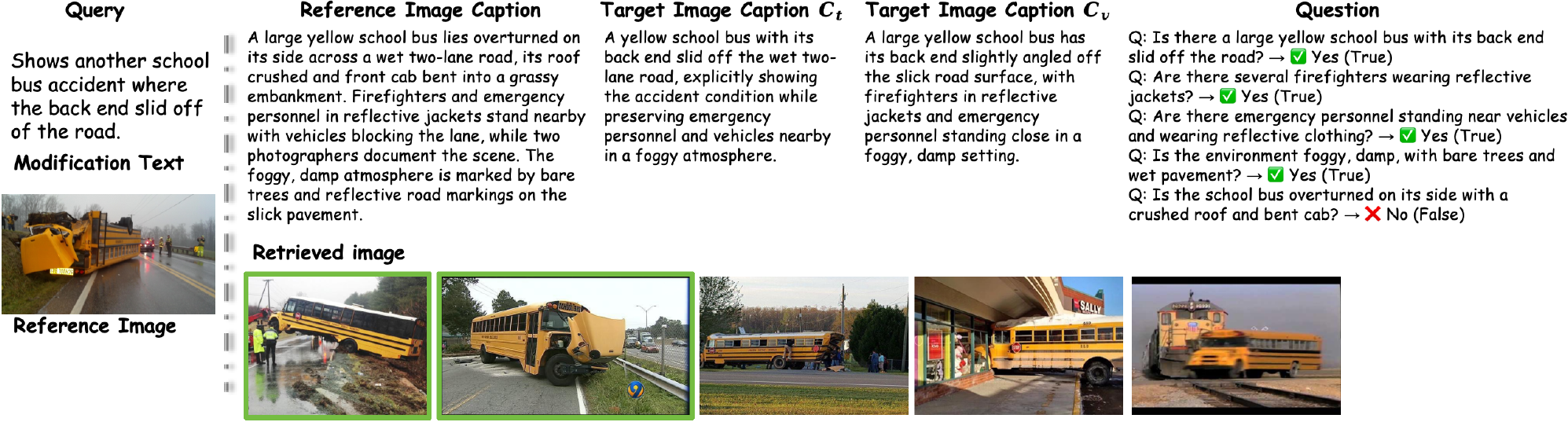}
  \caption{Case study on CIRR. $\text{X}^\text{R}$ correctly grounds complex scene edits (e.g., bus orientation, reflective jackets) through factual verification. Target image is marked with the {\color{casegreen}green box}.}
  \label{fig:case_stduy}
\end{figure*}

\begin{figure*}[!ht]
  \centering
  \includegraphics[width=\linewidth]{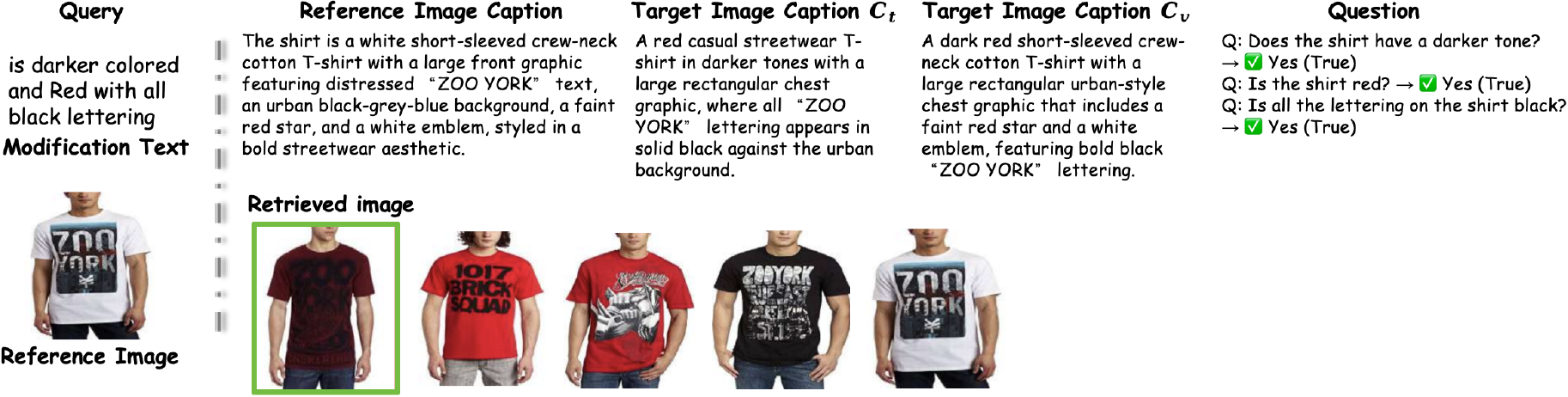}
  \caption{Case study on FashionIQ. $\text{X}^\text{R}$ captures subtle attribute edits (e.g., tone, lettering) and validates them via text-based questioning. Target image is marked with the {\color{casegreen}green box}.}
  \label{fig:case_stduy1}
\end{figure*}

\begin{figure*}[!ht]
  \centering
  \includegraphics[width=\linewidth]{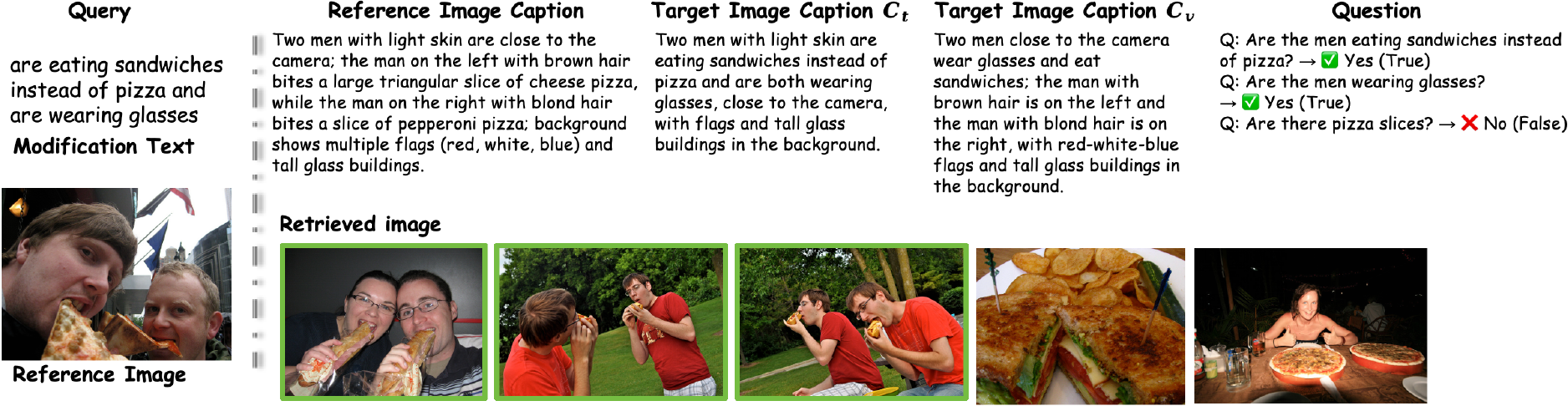}
  \caption{Case study on CIRCO. $\text{X}^\text{R}$ remains robust under distractor-heavy settings by verifying entity-level edits (e.g., food type, clothing). Target image is marked with the {\color{casegreen}green box}.}
  \label{fig:case_stduy2}
\end{figure*}

Beyond aggregate metrics, we present case studies on CIRR, FashionIQ, and CIRCO to illustrate how $\text{X}^\text{R}$ behaves on concrete queries. 
These examples highlight complementary aspects of the framework: on CIRR, it grounds complex scene edits through factual verification; on FashionIQ, it captures subtle attribute modifications such as color or lettering; and on CIRCO, it remains robust under distractor-heavy settings where static similarity often fails. 
Together, these cases not only showcase the strengths of multi-agent reasoning but also reveal remaining challenges, offering qualitative evidence that complements our quantitative findings.

\end{document}